\documentclass[a4paper,12pt]{article}
\pdfoutput=1
\usepackage{bbm}
\usepackage{amsfonts,amsthm,amsmath,amssymb,latexsym,graphics,braket,color,ascmac,subfigure}
\usepackage[dvipdfm]{graphicx}
\usepackage{bm}
\usepackage{cite}
\usepackage[backref=page]{hyperref}

\numberwithin{equation}{section}
\begin{document}
\begin{titlepage}
\begin{flushright}
{\small \today}
 \\
\end{flushright}

\begin{center}

\vspace{1cm}

\hspace{3mm}{\bf \Large Black hole as a multipartite entangler:\vspace{2mm}\\
multi-entropy in AdS${}_3$/CFT${}_2$} \\[3pt] 

\vspace{1cm}

\renewcommand\thefootnote{\mbox{$\fnsymbol{footnote}$}}
Takanori {Anegawa}${}^{1}$\footnote{takanegawa@gmail.com},
Shota {Suzuki}${}^{2}$\footnote{
chso25001@g.nihon-u.ac.jp}, and
Kotaro {Tamaoka}${}^{2}$\footnote{tamaoka.kotaro@nihon-u.ac.jp}\\
\vspace{5mm}

${}^{1}${\small \sl Yonago College, National Institute of Technology
Yonago, Tottori 683-8502, Japan}\\
${}^{2}${\small \sl Department of Physics, College of Humanities and Sciences, Nihon University, \\Sakura-josui, Tokyo 156-8550, Japan}\\

\end{center}

\vspace{5mm}

\noindent
\abstract
We study multipartite entanglement in typical pure states holographically dual to pure BTZ black holes, using multi-entropy and its ``genuine'' version. In the bulk, these quantities are computed by minimal geodesic networks (so-called Steiner trees). We find that at sufficiently high temperature, the genuine tripartite multi-entropy exhibits a volume-law scaling in sharp contrast to vacuum AdS$_3$, where the genuine contribution is universal and size-independent. Moreover, we find another phase: once one subsystem exceeds half of the total system, the leading genuine tripartite entanglement vanishes and reduces to that for global AdS${}_3$. This transition is indeed consistent with recent arguments for distillable EPR pairs in tripartite Haar-random states. Motivated by finite-cutoff holography, we further study the radial cutoff dependence of multi-entropy and show that genuine multi-entropy acquires nontrivial size dependence even for the tripartite case in AdS${}_3$. As a byproduct, we also observe an intriguing ``area-law'' contribution to multi-entropy that is relevant to vacuum AdS${}_3$.

\end{titlepage}
\setcounter{footnote}{0}
\renewcommand\thefootnote{\mbox{\arabic{footnote}}}
%%%%%%%%%%%%%%%TITLE%%%%%%%%%%%%%%%%\newpage
\tableofcontents
\flushbottom
\section{Introduction and Summary}

The AdS/CFT correspondence~\cite{Maldacena:1997re} has long suggested a deep connection between bulk geometry and quantum entanglement in the boundary theory. In particular, the Ryu-Takayanagi (RT) formula~\cite{Ryu:2006bv} and its generalizations~\cite{Hubeny:2007xt,Faulkner:2013ana,Engelhardt:2014gca} indicate that spacetime geometry is, to a remarkable extent, encoded in the pattern of entanglement across boundary subregions. As an interesting development, this perspective has motivated tensor network interpretations of the AdS/CFT correspondence~\cite{Swingle:2009bg}, in which spacetime is represented as a network of tensors whose contraction pattern reproduces boundary entanglement. Concrete realizations of this idea include MERA~\cite{Vidal:2006sxo}, its continuum variants~\cite{Haegeman:2011uy}, and quantum error-correcting codes~\cite{Pastawski:2015qua}.

A crucial feature underlying these constructions is the existence of isometries of AdS spacetime. The presence of the spacetime isometry naturally leads to tensor networks built from identical patterns of tensors, reflecting the isometry of the background geometry. In this sense, vacuum AdS admits a particularly simple entanglement structure: the same local entangling pattern is repeated throughout the bulk.

The situation changes qualitatively once an energy excitation corresponding to a black hole is introduced on each side. The presence of a horizon and an interior region breaks the global homogeneity of spacetime, and it is no longer obvious how a tensor network description should be organized. In particular, the degrees of freedom associated with the black hole cannot be captured by a naive repetition of the same tensor used in vacuum AdS. This raises a natural question: what kind of entanglement structure characterizes the tensors associated with black holes? Answering this question is essential for understanding black holes themselves and the quantum information theoretic approach to holography. 

A natural diagnostic of such structures is multipartite entanglement, namely, entanglement shared among multiple subsystems. While bipartite entanglement has long been regarded as playing a central role in AdS/CFT, it has become increasingly evident in recent years that multipartite entanglement also plays a crucial role~\cite{Umemoto:2019jlz, Akers:2019gcv,Hayden:2021gno} (see also recent discussions~\cite{Iizuka:2025bcc, Balasubramanian:2025hxg}).

A class of quantities known as multi-entropy~\cite{Gadde:2022cqi,Penington:2022dhr} has been proposed as a systematic way to quantify multipartite entanglement for pure states. By construction, the ${\mathtt q}$-th multi-entropy also counts entanglement involving lower than ${\mathtt q}$-parties. After subtracting all lower-partite contributions, one obtains the so-called genuine multi-entropy~\cite{Penington:2022dhr,Iizuka:2025ioc}, which is expected to capture only irreducible multipartite entanglement. In the tripartite case, this subtraction is unique and unambiguous, while for four or more parties residual ambiguities remain, signaling the need for further physical input to fully characterize higher-partite entanglement~\cite{Iizuka:2025ioc}. Refer also to recent developments~\cite{Gadde:2023zzj, Gadde:2023zni, Harper:2024ker, Liu:2024ulq, Yuan:2024yfg, Gadde:2024taa, Iizuka:2024pzm, Iizuka:2025caq, Mori:2025gqe, Harper:2025uui, Gadde:2025csh, Iizuka:2025elr, Berthiere:2025toi, Yuan:2025dgx, Ahn:2025bdm, Iizuka:2025pqq, Sheffer:2025zyr}.

In this work, we investigate how multipartite entanglement, especially tripartite entanglement, is realized in holographic states dual to AdS black holes, and how it differs from the vacuum AdS case. Our central goal is to clarify to what extent black holes act as ``multipartite entanglers,’’ and how much genuine multipartite entanglement they can support. Since multi-entropy is defined only for pure states, we focus on pure black hole states, dual to typical high-energy pure states in holographic CFTs. Focusing on AdS${}_3$/CFT${}_2$, we compute holographic multi-entropy using its geometric dual description in terms of minimal Steiner-tree-like surfaces in the bulk. For the static BTZ black hole, we find that at sufficiently high temperature, genuine multipartite entanglement exhibits a volume-law–type behavior, scaling linearly with the sizes of boundary subregions. This sharply contrasts with the vacuum AdS case, where the genuine tripartite entanglement remains universal and independent of subsystem size~\footnote{A similar analysis of holographic multi-entropy was discussed in~\cite{Harper:2024ker} from the perspective of consistency with 2D CFTs in a particular phase. In the present paper, we extend the analysis to a more general setting and find the existence of nontrivial phase transitions.}.

A particularly striking feature emerges in the tripartite case: when at least one of the three boundary subsystems becomes larger than half of the total system, the genuine tripartite entanglement vanishes at leading order and settles down to that of vacuum AdS. The genuine tripartite entanglement reaches its maximum when the three subsystems have equal sizes, and the leading order contribution is proportional to the Bekenstein–Hawking entropy. This behavior is in agreement with a recent argument using tripartite Haar-random states~\cite{Li:2025nxv}. By considering disconnected subsystems, we also observe, as a byproduct, an ``area-law'' contribution to multi-entropy that is relevant to vacuum AdS${}_3$.

We further extend our analysis to include finite radial cutoffs in the bulk. Introducing a finite cutoff allows us to probe how multipartite entanglement evolves from the UV to the IR, and provides a natural connection to $T\overline{T}$-deformations~\cite{Zamolodchikov:2004ce,Smirnov:2016lqw} in AdS/CFT~\cite{McGough:2016lol}. We show that, in contrast to the universal result in global AdS${}_3$, genuine multi-entropy acquires a nontrivial size dependence at finite cutoff and decreases monotonically as the cutoff is lowered. This observation suggests that the universal constant contribution observed in the UV originates from conformal symmetry, while deeper in the bulk, multipartite entanglement is distributed across multiple scales even for the ground state of holographic CFTs. We also find that this monotonic behavior becomes milder for pure BTZ black holes, suggesting that the effect of backreaction enhances multipartite entanglement as well.

Taken together, our results demonstrate that black holes dramatically enhance multipartite entanglement compared to global AdS and imprint characteristic signatures—such as volume-law growth and phase transitions—on genuine multi-entropy. These findings provide concrete constraints on the structure of ``black-hole tensors’’ in holographic tensor networks.

The organization of the rest of this paper is as follows. In Section~\ref{sec:review}, we review the definition of (genuine) multi-entropy in multipartite quantum systems, together with their holographic dual descriptions. In Section~\ref{sec:emb}, we develop a systematic method for computing holographic multi-entropy using the embedding space formalism and formulate the problem in terms of minimal Steiner trees in the bulk. 
Section~\ref{sec:BTZ} is devoted to the analysis of multipartite entanglement in the
static BTZ black hole, where we clarify the phase structure of genuine multi-entropy
and demonstrate its characteristic behavior and transitions. 
In Section~\ref{sec:finitecutoff}, we study the effects of introducing a finite radial
cutoff, discuss the resulting scale dependence of multipartite entanglement. 
Finally, in Section~\ref{sec:discussion}, we discuss the implications of our results and outline future directions.

\paragraph{Note added:}After the completion of this manuscript, we noticed that a related paper~\cite{Ju:2025eyn} appears, which has some overlap with our discussion in Section~\ref{sec:finitecutoff}.

\section{Review of multipartite quantities}\label{sec:review}
In this section, we review the definition of multi-entropy and its holographic dual in multipartite systems. Readers familiar with the subject may skip this section. This quantity is defined in \cite{Gadde:2022cqi,Penington:2022dhr} as a symmetric multipartite entanglement measure in $\mathtt q$-partite systems. 

In this paper, our main focus is only on multi-entropy in tripartite ($\mathtt q=3$) and fourpartite ($\mathtt q=4$) systems. We first introduce the general definition of multi-entropy for ${\mathtt q}$-partite systems, and then discuss in detail the specific cases ${\mathtt q}= 3$ and $4$, which are the main quantities of interest in this paper.

\subsection{Multi-entropy in multipartite systems}
Let us first consider the entanglement entropy in the case where the system can be divided into two subsystems $A_1$ and $A_2$, {\it i.e.,}\ the bipartite case.
Denoting the Hilbert space and the basis of $A_{\mathtt a}$ as $\mathcal{H}_{\mathtt a}$ and $\{ \ket{\alpha_{\mathtt a}}\}$, a general bipartite pure state can be written as $\ket{\Psi_2} = \sum_{\alpha_1} \sum_{\alpha_2} \psi_{\alpha_1 \alpha_2} \ket{\alpha_1}\otimes  \ket{\alpha_2}$. Tracing out $A_2$, we define the reduced density matrix of $A_1$ by $\rho_1 = {\rm Tr}_{A_2} \rho$, and the entanglement entropy is then defined by $S(A_1) = -{\rm Tr} \rho_1 \log \rho_1$. 
In practice, this computation is often carried out using the replica trick. For this purpose, the entropy is first rewritten in terms of the R\'enyi entropy $S_n(A_1)$ as follows.
\begin{align}
S_n(A_1) = \frac{1}{1-n}\log {\rm Tr} \rho_1^n\ ,\ \ \ \lim_{n \to 1} S_n(A_1) = S(A_1)
\end{align}
Therefore, the computation of the entropy is reduced to evaluating ${\rm Tr}\rho_1^n$. To compute this, one needs to prepare $n$ copies of $A_1$ and likewise $n$ copies of $A_2$. We label each copy by an index $i=1,\dots,n$, and denote the basis of the $i$-th copies as $\{\ket{\alpha_1^{(i)}}\}$ and $\{\ket{\alpha_2^{(i)}}\}$ respectively.
Writing the matrix elements of the reduced density matrix as $(\rho_1)^{\alpha_1^{(2)}}_{\alpha_1^{(1)}} = \psi_{\alpha_1^{(1)} \alpha_2^{(1)}} \bar{\psi}^{\alpha_1^{(2)} \alpha_2^{(1)}}$, then ${\rm Tr}\rho_1^n$ can be expressed as follows.
\begin{align}
{\rm Tr}\rho_1^n &= (\rho_1)^{\alpha_1^{(2)}}_{\alpha_1^{(1)}}(\rho_1)^{\alpha_1^{(3)}}_{\alpha_1^{(2)}} \cdots (\rho_1)^{\alpha_1^{(1)}}_{\alpha_1^{(n)}} = (\psi_{\alpha_1^{(1)} \alpha_2^{(1)}}\psi_{\alpha_1^{(2)} \alpha_2^{(2)}} \cdots \psi_{\alpha_1^{(n)} \alpha_2^{(n)}})(\bar{\psi}^{\alpha_1^{(2)} \alpha_2^{(1)}}\bar{\psi}^{\alpha_1^{(3)} \alpha_2^{(2)}} \cdots \bar{\psi}^{\alpha_1^{(1)} \alpha_2^{(n)}}) \notag \\
&= (\psi_{\alpha_1^{(1)} \alpha_2^{(1)}}\psi_{\alpha_1^{(2)} \alpha_2^{(2)}} \cdots \psi_{\alpha_1^{(n)} \alpha_2^{(n)}})(\bar{\psi}^{\alpha_1^{(\sigma_1 \cdot 1)} \alpha_2^{(\text{id}\cdot 1)}}\bar{\psi}^{\alpha_1^{(\sigma_1 \cdot 2)} \alpha_2^{(\text{id}\cdot 2)}} \cdots \bar{\psi}^{\alpha_1^{(\sigma_1 \cdot n)} \alpha_2^{(\text{id}\cdot n)}})
\end{align}
Here, in the second equality, the transitions among the replicas of $A_1$ are represented by the permutation $\sigma_1 = (1\,2\,\cdots\,n) \in \mathbb S^n$ which is the cyclic permutation of order $n$. In the same manner, the transitions among the replicas of $A_2$ are described by $\sigma_2 = \mathrm{id}$.
This corresponds to the fact that we first traced out $A_2$ in order to construct the reduced density matrix. 

Here we pause to summarize an important property of the entanglement entropy in bipartite systems. This quantity satisfies $S(A_1) = S(A_2)$ for pure state. In this sense, the entanglement entropy is symmetric under the exchange of $A_1$ and $A_2$. 

In fact, even in the multipartite case, an entanglement measure in which the replica transitions of one subsystem are given by \text{id} ({\it i.e.},\ tracing out), while those of the remaining subsystems are given by cyclic permutations of order $n$, enjoys a symmetry property analogous to that of the bipartite case: it is symmetric under exchanging the subsystems. See \cite{Gadde:2022cqi,Penington:2022dhr} for details. 
From the above, when we prepare the ${\mathtt q}$-partite pure state 
\begin{align}
\ket{\Psi_{\mathtt q}} = \sum_{\alpha_1}^{d_1} \cdots \sum_{\alpha_{\mathtt q}}^{d_{\mathtt q}} \psi_{\alpha_1 \cdots \alpha_{\mathtt q}} \ket{\alpha_1}\otimes \cdots \otimes \ket{\alpha_{\mathtt q}}\in \bigotimes_{{\mathtt a}=1}^{\mathtt q} {\mathcal{H}}_{\mathtt a},
\end{align}
we can define the following quantity
\begin{align}
\label{defZn}
Z_n^{(\mathtt q)} = (\psi_{\alpha_1^{(1)} \cdots \alpha_{\mathtt q}^{(1)}}\cdots \psi_{\alpha_1^{(n)} \cdots \alpha_{\mathtt q}^{(n)}})(\bar{\psi}^{\alpha_1^{(\sigma_1 \cdot 1)} \cdots \alpha_{\mathtt q}^{(\sigma_{\mathtt q} \cdot 1)}} \cdots \bar{\psi}^{\alpha_1^{(\sigma_1 \cdot n)} \cdots \alpha_{\mathtt q}^{(\sigma_{\mathtt q} \cdot n)}}).
\end{align}
This is the multipartite analogue of ${\rm Tr} \rho_1^n$ in the bipartite case. Here, among the permutations $\sigma_{\mathtt a}$ (${\mathtt a}=1,\ldots,{\mathtt q}$), one of them is taken to be the id.
The remaining permutations are taken to be the cyclic permutation of order $n$, which cyclically connects the replica indices of the ${\mathtt a}$-th subsystem.
Note that the choice of which $\sigma_{\mathtt a}$ is taken to be the identity is completely arbitrary. This does not amount to anything more than an appropriate relabeling of the replica indices. 

Using this, we compute the {\bf $n$-th R\'enyi multi-entropy} $S_n^{({\mathtt q})} (A_1 : A_2 : \cdots : A_{\mathtt q})$ and then take the $n\to 1$ limit appropriately, thereby defining a symmetric and natural {\bf multi-entropy} $S^{({\mathtt q})}(A_1 : A_2 : \cdots : A_{\mathtt q})$ for $\mathtt q$-partite systems:
\begin{align}
S_n^{({\mathtt q})} (A_1 : A_2 : \cdots : A_{\mathtt q}) &= \frac{1}{1-n}\frac{1}{n^{\mathtt q-2}} \log \frac{Z_n^{({\mathtt q})}}{(Z_1^{({\mathtt q})})^{n^{{\mathtt q}-1}}},\\
S^{({\mathtt q})}(A_1 : A_2 : \cdots : A_{\mathtt q}) &= \lim_{n \to 1} S_n^{({\mathtt q})} (A_1 : A_2 : \cdots : A_{\mathtt q}).
\end{align}
In the arguments of $S^{(q)}_n$ and $S^{(q)}$, the ordering of the parties is chosen so that the party corresponding to the identity permutation appears last. Accordingly, in the above definition, we take $\sigma_{\mathtt q} = \mathrm{id}$. For the remaining parties, the permutations are taken symmetrically, so their ordering is irrelevant. As discussed above, one can indeed show that this construction satisfies the property of being symmetric with respect to each party.
\begin{align}
S^{({\mathtt q})}( \cdots : A_m:\cdots :A_n :\cdots ) &= S^{({\mathtt q})}(\cdots : A_n:\cdots :A_m :\cdots )
\end{align}
For ${\mathtt q} = 2$, this reduces to the conventional entanglement entropy. In this case, $S^{(2)}(A_1 : A_2)$ (where the permutation associated with $A_2$ is taken to be $\sigma_2 = \mathrm{id}$, {\it i.e.},\ the trace is taken over $A_2$) will simply be denoted by $S(A_1) = S(A_2) \equiv S^{(2)}(A_1 : A_2).$

We now describe the geometric (not holographic) interpretation of the cases ${\mathtt q}=3$ and ${\mathtt q}=4$, which are particularly relevant to this paper. We begin with the case ${\mathtt q}=3$. For the pure state $\ket{\Psi_3} \in {\mathcal{H}}_A \otimes {\mathcal{H}}_B  \otimes {\mathcal{H}}_C$ in tripartite systems $ABC$, we can construct the reduced density matrices
\begin{align}
\rho \equiv \underset{C}{\rm Tr}\ \rho_{\rm tot}
\end{align}
using $\rho_{\rm tot} = \ket{\Psi_3}\bra{\Psi_3}$. This corresponds to taking $\sigma_3 = \mathrm{id}$. In this case, (\ref{defZn}) simply amounts to summing over the replica indices while transporting the indices of subsystems $A$ and $B$ according to $\sigma_1$ and $\sigma_2$, respectively.
Since $\sigma_1$ and $\sigma_2$ are both cyclic permutations of order $n$, the following interpretation is possible.
First, the object $\rho$ carries the index structure of $\rho_{i_A i_B}^{j_A j_B}$, and its diagrammatic representation is given as a point with four lines extending from it in the four directions. 
\begin{equation}
\rho_{i_A i_B}^{j_A j_B} =
 \vcenter{\hbox{\includegraphics[clip,scale=0.3]{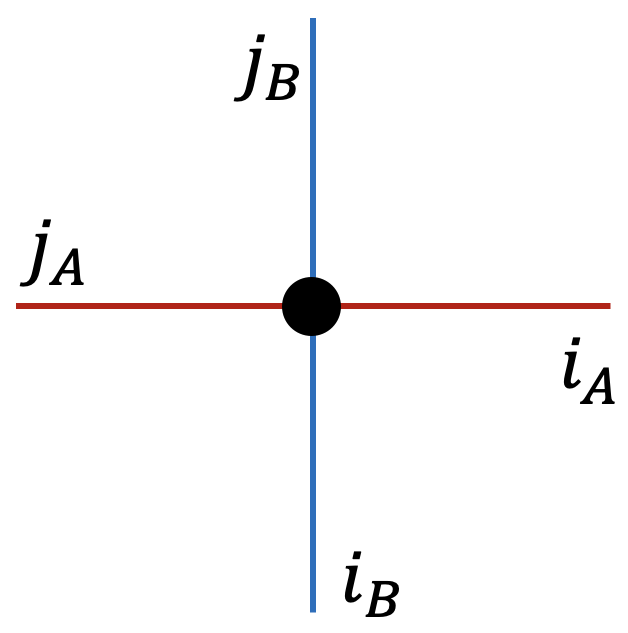}}}
\end{equation}
We then prepare $n^2$ replica copies of this object and arrange them on a square lattice. After that, from each $\rho$ placed on the lattice, we extend the lines both vertically and horizontally, interpret these connections as the diagrammatic legs of $\rho$, and perform contractions along every such line. In addition, periodic boundary conditions are imposed on the edges of the square lattice. For example, a line extending downward from the bottommost lattice site is identified with the line entering the topmost lattice site from above. The quantity obtained through this construction is defined as $Z_n^{(3)}$. 
This construction provides an intuitive lattice interpretation of the replica contractions. Interpreted as a diagram, 
\begin{equation}
Z_2^{(3)}=
 \vcenter{\hbox{\includegraphics[clip,scale=0.45]{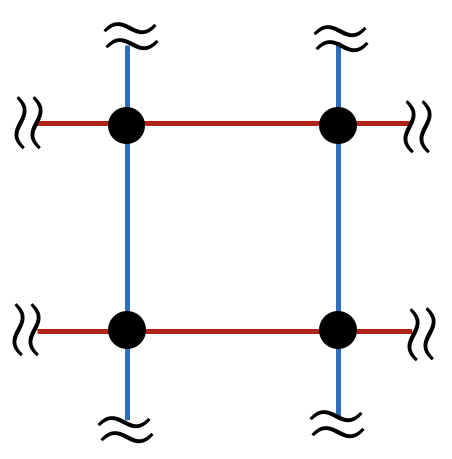}}}\ \ \ \ \  
Z_n^{(3)}=
 \vcenter{\hbox{\includegraphics[clip,scale=0.3]{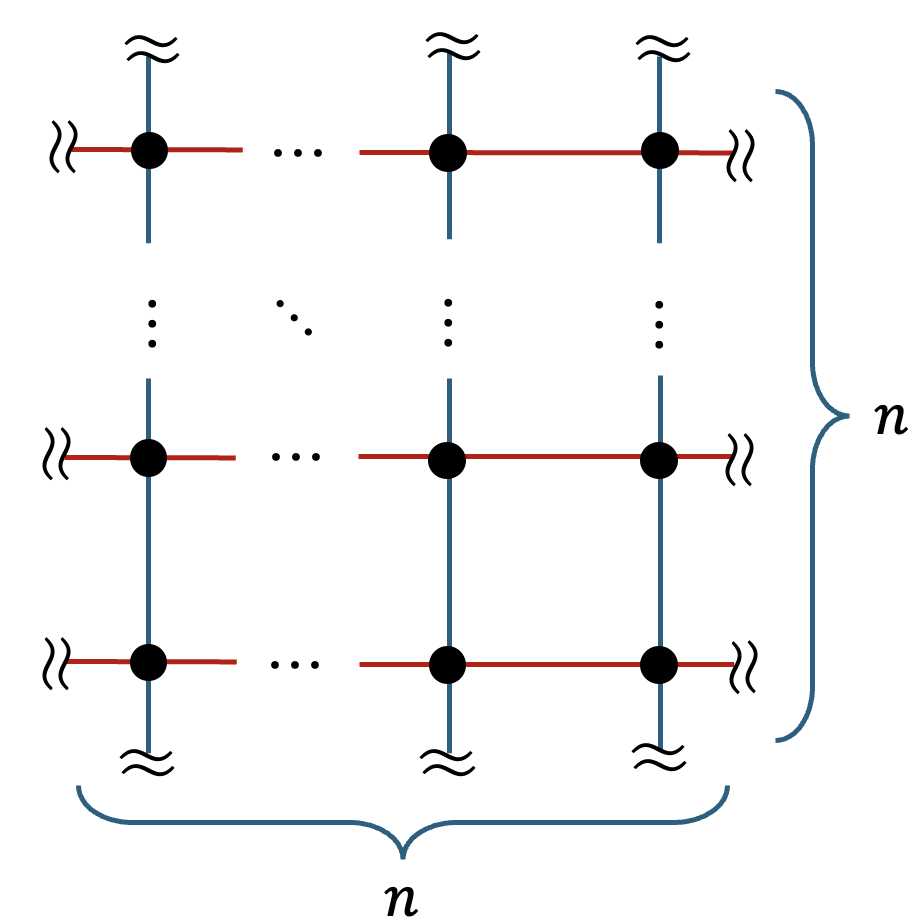}}}
\end{equation}

Similarly, we can find the diagrammatic representation for fourpartite systems. For the pure state $\ket{\Psi_4} \in {\mathcal{H}}_A \otimes {\mathcal{H}}_B  \otimes {\mathcal{H}}_C \otimes \mathcal{H}_D$ in a fourpartite systems ABCD, we can construct the reduced density matrices in the same way $\rho \equiv \underset{D}{\rm Tr}\ \rho_{\rm tot}$ from $\rho_{\rm tot} = \ket{\Psi}\bra{\Psi}$. The object $\rho$ carries the index structure of $\rho_{i_A i_B i_C}^{j_A j_B j_C}$, and it is understood diagrammatically as shown below.
\begin{equation}
\rho_{i_A i_B i_C}^{j_A j_B j_C}=
 \vcenter{\hbox{\includegraphics[clip,scale=0.35]{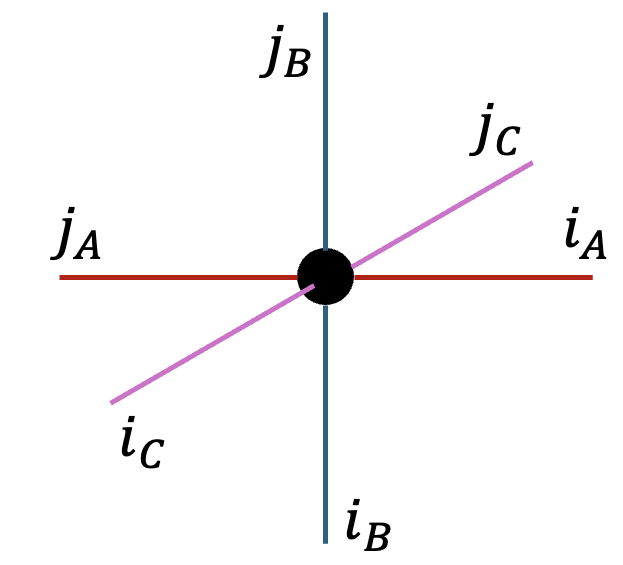}}},
\end{equation}
We now prepare $n^3$ replica copies of $\rho$ and arrange them in a cubic $n \times n \times n$ lattice. As before, we extend lines from each copy in the vertical, horizontal, and depth directions, interpret these lines as the legs of the diagrammatic representation of $\rho$, and contract all of them accordingly. Periodic boundary conditions are imposed in all directions. For example, a line extending downward from a site on the bottom face of the cube is identified with the line entering the corresponding site on the top face. The resulting quantity obtained through this construction is defined as $Z_n^{(4)}$. Although drawing the full diagram is cumbersome, the case of $Z_2^{(4)}$ may be represented schematically as follows.
\begin{equation}
Z_2^{(4)}=
 \vcenter{\hbox{\includegraphics[clip,scale=0.45]{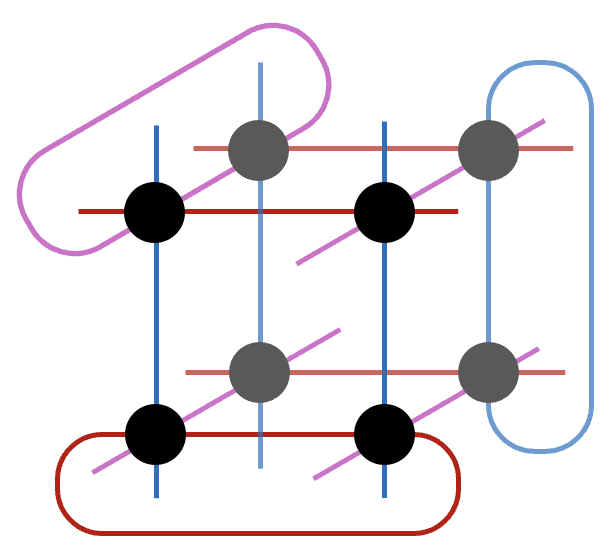}}}
\end{equation}

In general, the multi-entropy $S^{(\mathtt q)}$ contains not only the contribution from genuine $\mathtt q$-partite entanglement, but also those arising from its subsystems, namely the $\mathtt k$-partite entanglement with $ \mathtt k < \mathtt q$. Even for a state composed purely of bipartite entanglement, such as $\ket{{\rm PB}}_{ABC}=|\psi_1\rangle_{A_1B_1} \otimes |\psi_2\rangle_{A_2C_1}\otimes |\psi_3\rangle_{B_2C_2}$, the multi-entropy generally returns a non-zero value.  
In particular, one finds
\begin{align}
S^{(3)}(A:B:C)= \frac{1}{2}(S(A)+S(B)+S(C)).
\end{align}
Therefore, to extract only the contribution of genuine $\mathtt q$-partite entanglement, we define the {\bf genuine multi-entropy} by subtracting all lower-partite contributions \cite{Liu:2024ulq,Iizuka:2025ioc}\footnote{In particular, we adopt the notation used in (5.3) and (5.4) of \cite{Iizuka:2025ioc}.}.
In particular, for \(\mathtt q = 3,4\), it is defined as follows.
\begin{align}
{\rm GM}^{(3)}&(A:B:C)= S^{(3)}(A:B:C) - \frac{1}{2}(S(A)+S(B)+S(C))\\
{\rm GM}^{(4)}&(A:B:C:D) = S^{(4)}(A:B:C:D) \notag \\
&\ \ - \frac{1}{3}\left({\rm GM}^{(3)}(AB:C:D) +{\rm GM}^{(3)}(AC:B:D) + {\rm GM}^{(3)}(AD:B:C) \right. \notag \\
&\ \ \ \ \left. +{\rm GM}^{(3)}(BC:A:D)+ {\rm GM}^{(3)}(BD:A:C)+{\rm GM}^{(3)}(CD:A:B)\right) \notag \\
&\ \ + \left(a-\frac{1}{3}\right)(S(AB)+S(AC)+S(AD)) - \left( a+\frac{1}{6} \right)(S(A)+S(B)+S(C)+S(D))
\label{GM4def}
\end{align}
In fact, the state \(\ket{{\rm PB}}_{ABC}\) above does not contain any genuine tripartite entanglement, and indeed one finds that ${\rm GM}^{(3)}(A:B:C) = 0$. It should be noted that, in the four-partite case, the quantity ${\rm GM}^{(4)}(A:B:C:D)$ contains a free parameter $a$ that cannot be fixed solely from the conditions discussed above, reflecting an intrinsic ambiguity in characterizing genuine four-partite entanglement.

\subsection{Holographic dual of multi-entropy}
The multi-entropy serves as a natural definition of entanglement in multipartite quantum systems, and, much like the RT surface, the holographic correspondence should be understood as well. In what follows, we review the holographic dual proposed in \cite{Gadde:2022cqi,Penington:2022dhr,Gadde:2023zzj}. The main objective of this paper is to understand the multi-entropy through this prescription~\footnote{While some analyses are also justified from the CFT side~\cite{Harper:2024ker} and tensor network models, we have to note that we cannot naively justify the proposed holographic dual using Lewkowycz-Maldacena type argument as analytic continuation for $n>2$ fails due to the replica symmetry breaking~\cite{Penington:2022dhr}. See also \cite{Gadde:2024taa}.}.

We assume the AdS$_3$/CFT$_2$ correspondence, where the boundary field theory is defined on $ \mathbb{R} \times S^1 $. Throughout this paper, we restrict our discussion to static spacetimes, so that
the relevant quantities can be defined on a single time slice. On a chosen time slice, we partition the boundary circle $S^1$ into ${\mathtt q}$ subsystems $\{ A_i \}$ ($i = 1, \ldots, {\mathtt q}$) and compute the ${\mathtt q}$-party multi-entropy. In AdS$_3$, the dual geometric quantity is obtained as the area of a codimension-2 surface according to the following prescription.

\subsubsection*{Bulk prescription for multi-entropy}\label{subsec:review_hol}
\begin{enumerate}
\item Partition the bulk region into domains $\{ \mathcal{D}_i \}$, each homologous to $A_i$. Their intersection with the boundary time slice must not be empty. In other words, $\partial \mathcal{D}_i = A_i$ and, if $\mathcal{B}$ denotes the boundary time slice, then $\mathcal{D}_i \cap \mathcal{B} \neq \varnothing$ for each $i$. Note that each $\mathcal{D}_i$ is allowed to be disconnected.

\item Between every pair $\{ \mathcal{D}_i, \mathcal{D}_j\}$, there may exist a codimension-2 surface $\mathcal{W}_{ij}$ separating them (in some cases, such a surface does not appear). We denote the union of all such surfaces by $\mathcal{W}$.

\item The multi-entropy is given by minimizing the total area of $\mathcal{W}$ over all allowed partitions $\{ \mathcal{D}_i \}$. More precisely,
\begin{align}
S^{({\mathtt q})} (\{ A_i \}) = \frac{1}{4G_N}  \min_{ {\rm all}\ \{ \mathcal{D}_i \} } {\rm Area}(\mathcal{W}).
\end{align}
\end{enumerate}

In the following, we illustrate several examples of the surface $\mathcal{W}$ in the cases that constitute the main focus of this paper. As a simple setup, consider partitioning the boundary circle $S^1$ into three regions $A$, $B$, and $C$, and suppose we wish to compute the corresponding multi-entropy. The bulk domains $\mathcal{D}_A$, $\mathcal{D}_B$, and $\mathcal{D}_C$, each of which is bounded by and homologous to $A$, $B$, and $C$, respectively, may be depicted schematically as shown below. 

\begin{equation}
S^{(3)}(A:B:C) = \frac{1}{4G_N} \times \min_{\{ \mathcal{D}_i \} }\, {\rm Area}\left[
 \vcenter{\hbox{\includegraphics[clip,scale=0.4]{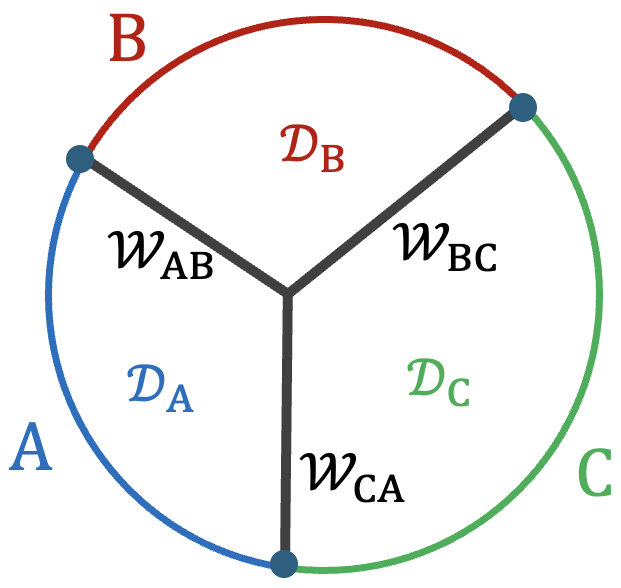}}}\right].\label{eq:hm1v}
\end{equation}

In particular, the surfaces $\mathcal{W}_{AB}$, $\mathcal{W}_{BC}$, and
$\mathcal{W}_{CA}$, which separate the respective domains $\{\mathcal{D}_i\}$,
together form a combined surface $\mathcal{W}$ (depicted by a black line). Since we search for the configuration with minimal area, $\mathcal{W}$ is assumed to be primarily composed of geodesic segments, and branching is allowed. Consequently, $\mathcal{W}$ takes the structure of what is commonly referred to as a Steiner tree. One then varies the position of the branching point to determine the configuration that minimizes the area of $\mathcal{W}$. A detailed analysis of this minimization procedure will be deferred to Section~\ref{sec:emb}.

For the genuine multi-entropy, there exists an important statement that can be established in holographic settings. For instance, in the tripartite case, one can show~\cite{Iizuka:2025ioc} that 
\begin{align}
{\rm GM}^{(3)}(A:B:C) \ge 0.
\end{align}
Since this inequality is not a universal property satisfied by generic quantum states, it indicates that the genuine multi-entropy captures a characteristic feature specific to holographic states.  
The proof proceeds in a manner analogous to the standard argument for RT surfaces such as~\cite{Headrick:2007km, Hayden:2011ag}, as outlined below:
\begin{align}
 \vcenter{\hbox{\includegraphics[clip,scale=0.5]{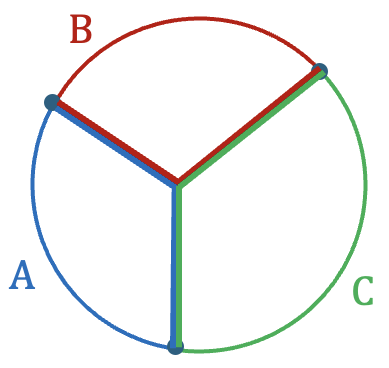}}} &\ge  \vcenter{\hbox{\includegraphics[clip,scale=0.5]{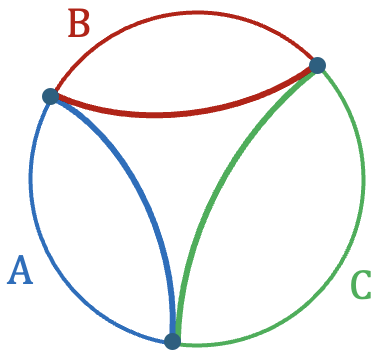}}}, \\
 2S^{(3)}(A:B:C)& \ge S(A)+S(B)+S(C).
\end{align}
The inequality holds because each surface on the right-hand side is, by
definition, the minimal surface with area no larger than that of the
corresponding kinked surface on the left-hand side (depicted by the same
color).

\begin{figure}[h]
 \centering
 \includegraphics[keepaspectratio, scale=0.5]{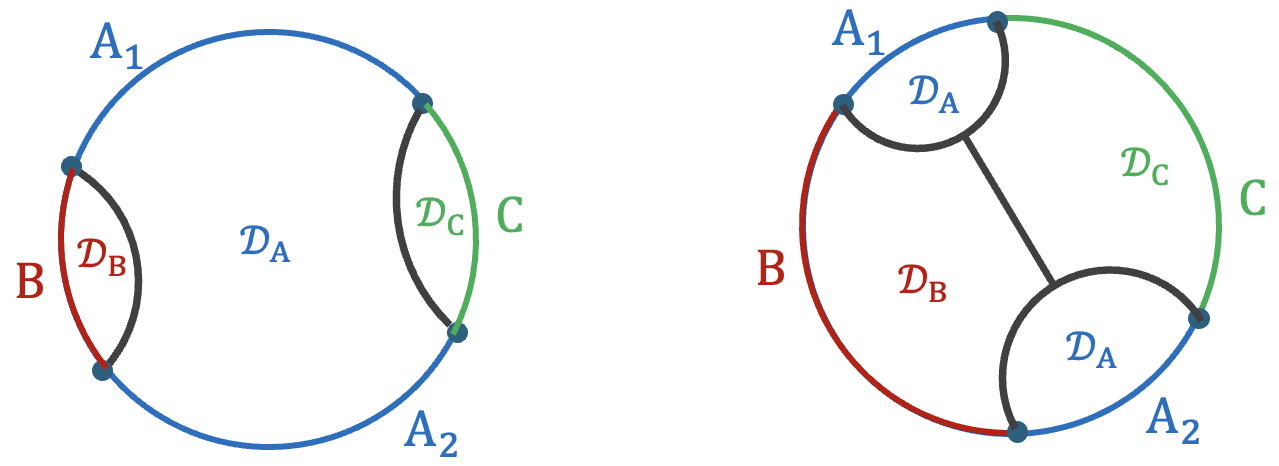}
 \caption{Bulk dual of $S^{(3)}(A:B:C)$ in the case where region $A$ is disconnected and composed of $A_1$ and $A_2$.  
Left: When $A_1$ and $A_2$ are sufficiently large compared with the total boundary region, $\mathcal{W}$ is given by the RT surfaces associated with regions $B$ and $C$.  
Right: When $A_1$ and $A_2$ are sufficiently small compared with the total boundary region, $\mathcal{W}$ consists of a set of geodesic segments that allow for two branch points.}
\label{fig:S3bulkdis}
\end{figure}

Even in the tripartite case, one may encounter situations in which some of the boundary regions are disconnected. As an illustrative example, let us consider the case in which region $A$ is disconnected, and denote its two components by $A_1$ and $A_2$. Depending on the relative sizes of $A_1$ and $A_2$, the bulk admits two possible patterns of domain partition. When the disconnected components are relatively large, the minimal-area configuration of $\mathcal{W}$ is simply given by the RT surfaces associated with regions $B$ and $C$. In this case, ${\rm GM}^{(3)}(A:B:C) = 0$ since $S^{(3)}(A:B:C) = S(B)+S(C)$. However, as the size of these components increases, the area of the RT surface
also grows, and beyond a certain size the configuration undergoes a transition
to a different phase. In this phase, $\mathcal{W}$ corresponds to a bulk
partition in which the relevant surfaces have two branching points. See
Figure~\ref{fig:S3bulkdis} for a schematic illustration.

The bulk dual for the fourpartite case can be constructed in a similar manner.  
In this case, the configuration with two branching points gives a smaller area than the one with a single branching point, and therefore the multi-entropy can be computed from the area of the corresponding surface $\mathcal{W}$.  

The figure in the equation below corresponds to the situation in which the sizes of regions $A$ and $B$ are relatively small, so that the minimal configuration of $\mathcal{W}$ takes the s-channel form.  
Depending on the relative sizes of the subsystems, however, the t-channel configuration may become the minimal one instead.  
\begin{align}
\label{fig:S4dual}
S^{(4)}(A:B:C:D) = \frac{1}{4G_N} \times {\rm Area} 
 \vcenter{\hbox{\includegraphics[clip,scale=0.35]{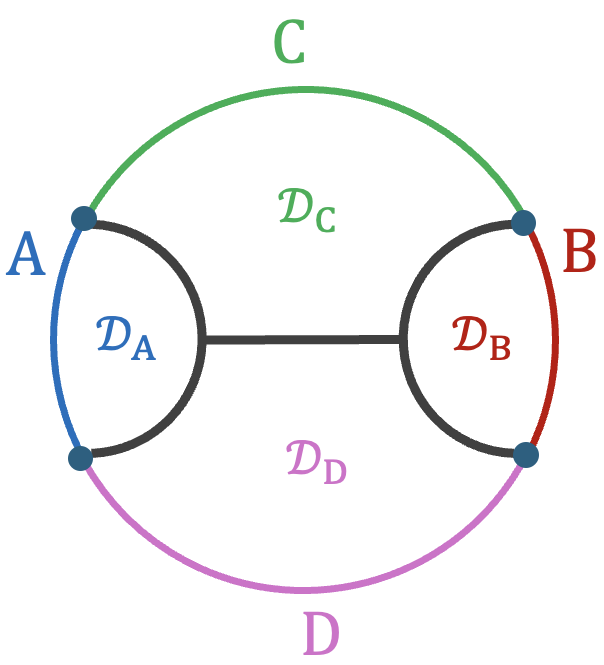}}}
\end{align}
Even in that case, the same statement $ \left. {\rm GM}^{(4)}(A:B:C:D)\right|_{a=1/3} \ge 0$ can be established~\cite{Iizuka:2025caq}.

The situation becomes more involved in the presence of a black hole in the bulk.
We discuss this case in detail in the remainder of this paper.

\section{Multi-entropy from embedding space formalism}\label{sec:emb}
In this section, we develop some useful formulas for calculating multi-entropy by using the embedding space formalism. The readers who are not interested in technical details may skip this section.  

We begin by introducing the embedding space formalism used throughout this paper.
AdS$_3$ spacetime and its quotients are represented as the hyperboloid inside $\mathbb{R}^{2,2}$,
\begin{align}
 X \cdot X \equiv\eta_{AB}X^AX^B\equiv -(X^0)^2 - (X^1)^2 + (X^2)^2 + (X^3)^2 = -\ell^2 ,
 \label{eq:ads-hyperboloid}
\end{align}
and we set $\ell = 1$ unless otherwise stated. 
Spacelike geodesic distance between points $X_1, X_2$ is given by
\begin{align}
 \sigma(X_1 , X_2) = \cosh^{-1}(- X_1 \cdot X_2).
 \label{eq:geodesic-embedding}
\end{align}
In particular, global AdS${}_3$ $(t,r,\varphi)$ can be embedded as
\begin{align}
 X^0 &= \sqrt{r^2+1}\cos t, \\
 X^1 &= \sqrt{r^2+1}\sin t, \\
 X^2 &= r \sin\varphi, \\
 X^3 &= r \cos\varphi .
 \label{eq:global-embed}
\end{align}
One can also express the BTZ black holes in terms of the embedding space. For the non-rotating BTZ black hole with horizon radius $r_{+}$, the embedding coordinates in the covering space may be written as
\begin{align}
 X^0 &= \dfrac{r}{r_+}\,\cosh(r_{+}\varphi), \\
 X^1 &= \dfrac{\sqrt{r^2 - r_{+}^2}}{r_+}\,\sinh(r_{+} t), \\
 X^2 &= \dfrac{r}{r_+}\,\sinh(r_{+}\varphi), \\
 X^3 &= \dfrac{\sqrt{r^2 - r_{+}^2}}{r_+}\,\cosh(r_{+} t).
 \label{eq:btz-embed}
\end{align}
Since we have the identification $\varphi\sim\varphi+2\pi$, we need to impose the identification 
\begin{align}
X^0\pm X^2\sim e^{2\pi r_+}(X^0\pm X^2) \label{eq:id_BTZ}
\end{align}
This clearly defines the BTZ black holes as the orbifold of AdS${}_3$. 

In what follows, using the embedding-space formalism, we study the multipartite extremal surfaces (called Steiner trees) anchored on ``boundary points'' $\{X_1,X_2,\cdots\}$,  sharing the same radius $r\equiv r_b$, with ``bulk vertices'' $\{Y_1, Y_2,\dots\}$.
A similar analysis was carried out in earlier work~\cite{Gadde:2023zzj}, while our treatment is more general and, in particular, applies naturally to finite-cutoff geometries as well as to BTZ black hole backgrounds.

We minimize an area functional defined in the embedding space as follows:
\begin{align}
L(Y,\Lambda)&=\sum_{i=1}^3\sigma(X_i,Y)+\Lambda(Y^2+1),\label{eq:area}
\end{align}
where $\sigma(X_i,Y)$ is spacelike geodesic distance defined in \eqref{eq:geodesic-embedding}. 
One may interpret the minimized area as Eq.~\eqref{eq:hm1v} for AdS$_3$, while for the BTZ black hole there is the caveat that multiple candidates for the minimal area exist, as discussed later.

To take advantage of the embedding space, we treat the bulk point $Y$ as unconstrained by the AdS condition $Y^2+1=0$, and instead impose this condition as a constraint through the second term in \eqref{eq:area}. In doing so, the minimality condition reduces to
\begin{align}
\frac{\partial L}{\partial Y^A}=\sum^3_{i=1}\frac{-X_{iA}}{\sqrt{(-X_i \cdot Y)^2-1}}+2\Lambda Y_{A}=0, \label{eq:eom1}
\end{align}
and
\begin{align}
\frac{\partial L}{\partial \Lambda}=Y^2+1=0.\label{eq:eom2}
\end{align}
Multiplying $Y^A$ with \eqref{eq:eom1}, we obtain the expression for the Lagrange multiplier $\Lambda$ as
\begin{align}
\Lambda=\frac{1}{2}\sum_{i=1}^3\frac{(-X_i\cdot Y)}{\sqrt{(-X_i\cdot Y)^2-1}}.
\end{align}
This gives rise to
\begin{align}
\sum_{i=1}^3\frac{1}{\sqrt{(-X_i\cdot Y)^2-1}}\left[X_{iA}+(X_i\cdot Y)Y_{A}\right]=0.
\end{align}
Contracting with $X_j\;(j=1,2,3)$, we obtain
\begin{align}
\sum_{i=1}^3\frac{(-X_i\cdot X_j)-(-X_i\cdot Y)(-X_j\cdot Y)}{\sqrt{(-X_i\cdot Y)^2-1}}=0\ \ \ \; (j=1,2,3). \label{eq:minimal}
\end{align}
For large $r\equiv r_b$ limit, which is relevant to UV cutoff $\epsilon^{-1}_{UV}$ in CFT, $X_i$ can be written as $X_i=r_b P_i+\mathcal{O}(r_b^0)$, where 
\begin{align}
    P_i^2=0.
\end{align}
In this limit, the above equation simplifies significantly, and we obtain a general solution
\begin{align}
Y_A = \frac{1}{\sqrt{6}}\left(\sqrt{\frac{P_{23}}{P_{12}P_{13}}} P_{1A} + \sqrt{\frac{P_{13}}{P_{12}P_{23}}} P_{2A} + \sqrt{\frac{P_{12}}{P_{13}P_{23}}} P_{3A} \right),
\end{align}
where $P_{ij}\equiv -P_i\cdot P_j$. 
This solution at large $r_b$ has been already obtained in \cite{Gadde:2023zzj}. In particular, for AdS${}_3$ spacetime, this result implies
\begin{align}
    S^{(3)}(A:B:C)=\frac{3}{4G_N}\log\left[\frac{2}{\sqrt{3}}\right]+\frac{1}{8G_N}\sum_{1\leq i<j\leq3}\log(-2r_{b}^2P_i\cdot P_j),
\end{align}
thus we obtain
\begin{align}
{\rm GM}^{(3)}(A:B:C)=\frac{3}{4G_N}\log\left[\frac{2}{\sqrt{3}}\right].
\end{align}
As suggested in later sections, the overall coefficients $3$ would be identified with the number of boundary points for subsystems $A, B,$ and $C$. This ``area law'' appears to be parallel to the one for bipartite entanglement in 2D CFT. 

For pure state black holes, however, we cannot conclude the same constant value because we have a series of candidates of minimal geodesics due to the identification \eqref{eq:id_BTZ}. We will discuss this point in detail in the upcoming section.

For general $r_b$, which is not necessarily large, we can still solve the above equations, {\it e.g.,} by using Mathematica, albeit the solutions are quite involved in general. For the symmetric case, {\it i.e.}, $-X_1\cdot X_2=-X_2\cdot X_3=-X_3\cdot X_1$, one can easily find the simple analytic solution for general $r_b$: 
\begin{align}
    -X_1\cdot Y=-X_2\cdot Y=-X_3\cdot Y=\sqrt{\frac{(-2X_1\cdot X_2)+1}{3}}\;\;\;(\text{symmetric}),
\end{align}
 which is consistent with the large $r_b$ limit. 

One can extend these arguments to Steiner trees with more than one vertex, which are relevant to higher-partite multi-entropy or disconnected intervals discussed in Section~\ref{subsec:BTZtrid_four}. We leave the case of two vertices to Appendix~\ref{app:twov}.

\section{Multi-entropy in BTZ black hole}\label{sec:BTZ}
Since multi-entropy is defined only for pure states, in this section we discuss the typical {\it pure} state (black hole microstate) relevant to thermal state dual to static BTZ black hole~\footnote{Note that these states are not heavy primary states in 2D holographic CFT, which are indeed atypical~\cite {Datta:2019jeo,Kusuki:2019rbk,Kusuki:2019evw}. }. Effectively, it means that minimal surfaces can pass through the black hole without violating the homology constraints. Such states are also discussed in~\cite{Bao:2017guc}, for example.

\subsection{tripartite (adjacent intervals)}\label{sec:BTZtri}

We begin by considering the tripartite subsystems $A=[\varphi_1=0,\varphi_2], B=[\varphi_2,\varphi_3]$, and $C=[\varphi_3, 2\pi]$, which are mutually adjacent. Since we consider the theory on a circle, for each $\varphi_i$, we have an identification $\varphi_i\sim\varphi_i+2\pi n_i$ with $n_i\in\mathbb{Z}$. Therefore, each $X_i$ discussed in section \ref{sec:emb} has implicit $n_i$-dependence:
\begin{align}
-2X_i^{(n_i)}\cdot X_j^{(n_j)}=\dfrac{4r_b^2}{r_+^2}\sinh^2\left[\frac{r_+}{2}(\varphi_{ij}+2\pi n_{ij})\right]
\end{align}
where $\varphi_{ij}\equiv\varphi_i-\varphi_j$ and $n_{ij}\equiv n_i-n_j\in\mathbb{Z}$. A solution of the geodesic equation does not necessarily correspond to a minimal geodesic; therefore, we must select the minimal one subject to appropriate conditions.

For bipartite case, we simply choose $n_{ij}$ such that $\log(-2X_i^{(n_i)}\cdot X_j^{(n_j)})$ takes the minimal value. For $|\varphi_{ij}|<\pi$, we need to choose $n_{ij}=0$, and for $\pi<|\varphi_{ij}|<2\pi$, $n_{ij}=\pm 1$. This gives rise to
\begin{align}
    S(A)=S(A^c)=\begin{cases}
    \dfrac{1}{4G_N}\log\left[\frac{4r_b^2}{r_+^2}\sinh^2\left(\frac{r_+}{2}|\varphi_{ij}|\right)\right]&(0<|\varphi_{ij}|<\pi),\\
    \dfrac{1}{4G_N}\log\left[\frac{4r_b^2}{r_+^2}\sinh^2\left(\frac{r_+}{2}(2\pi-|\varphi_{ij}|)\right)\right]&(\pi<|\varphi_{ij}|<2\pi),
    \end{cases}
\end{align}
where $A^c$ denotes the complement of $A$. Other choices of $n_{ij}$s correspond to geodesics winding around the black hole, which are obviously not the minimal ones. Note that we are now dealing with a pure state, so the black hole entropy does not enter as a consequence of the homology constraint.

\begin{figure}[t]
    \centering
\includegraphics[width=6cm]{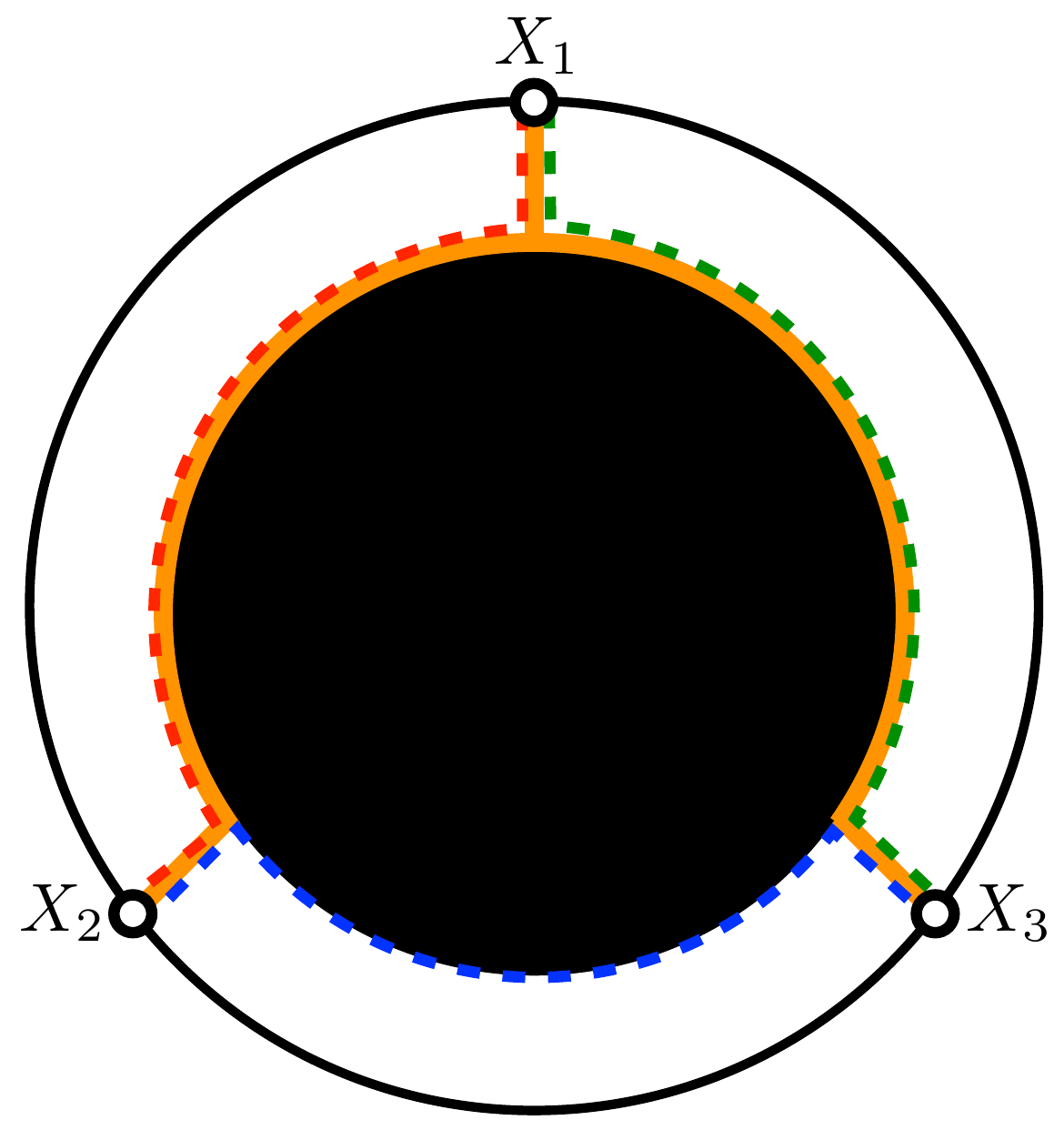}
    \caption{Schematic picture of the computation of the holographic tripartite multi-entropy. For later convenience, we take the large-$r_+$ limit here. The orange solid lines represent the Steiner tree $\mathcal{W}$ corresponding to the contribution to $S^{(3)}$. The dashed lines denote the RT surfaces used to compute the holographic entanglement entropy of each subsystem. Depending on the subsystem sizes, there are up to three distinct phases, characterized by which region ($\mathcal{D}_{A,B,C}$), separated by the tree $\mathcal{W}$, contains the black hole. The large contribution for the genuine tripartite multi-entropy ${\rm GM}^{(3)}$, proportional to the black hole entropy $S_{BH}$, arises from the difference between the paths of the Steiner tree and the RT surfaces. For the definition of $\mathcal{W}$ and $\mathcal{D}_{A,B,C}$, refer to Section \ref{subsec:review_hol}.}
    \label{fig:bh_larger}
\end{figure}

For the tripartite case, we need to minimize
\begin{align}
    S^{(3)}(A:B:C)=\frac{1}{4G_N}\min_{n_1,n_2,n_3\in\mathbb{Z}}\left[3\log\left[\frac{2}{\sqrt{3}}\right]+\frac{1}{2}\sum_{1\leq i<j\leq3}\log(-2X_i^{(n_i)}\cdot X_j^{(n_j)})\right],
    \label{eq:tri-minimize}
\end{align}
where
\begin{align}
-2X_1^{(n_1)}\cdot X_2^{(n_2)}&=\dfrac{4r_b^2}{r_+^2}\sinh^2\left[\frac{r_+}{2}(\varphi_{12}+2\pi n_{12})\right],\\
-2X_2^{(n_2)}\cdot X_3^{(n_3)}&=\dfrac{4r_b^2}{r_+^2}\sinh^2\left[\frac{r_+}{2}(\varphi_{23}+2\pi n_{23})\right],\\
-2X_3^{(n_3)}\cdot X_1^{(n_1)}&=\dfrac{4r_b^2}{r_+^2}\sinh^2\left[\frac{r_+}{2}(\varphi_{31}+2\pi n_{31})\right].
\end{align}
In particular, we cannot always choose $n_{i}$s so that each $-2X_i^{(n_i)}\cdot X_j^{(n_j)}$ can take the minimum value simultaneously. 

As an illustration, let us take $\varphi_1=0, \varphi_2=|A|, \varphi_3=|A|+|B|=2\pi-|C|$. Throughout this paper, we use $|X|$ as the total size of the subsystem $X$. We further assume $|B|=|C|=\pi-\frac{1}{2}|A|<\pi$ and take large-$r_+$ limit for the sake of presentation. In what follows, we would like to see $|A|$-dependence of $S^{(3)}$ and ${\rm GM}^{(3)}$.

In general, $S^{(3)}$ has three phases, depending on which domain the black hole contains. See Figure \ref{fig:bh_larger}. Since here we assumed $|B|=|C|$, we have only two phases depending on the size of $|A|$: 
\begin{align}
S^{(3)}(A:B:C)=\begin{cases}
\frac{3}{4G_N}\log\left[\frac{2}{\sqrt{3}}\right]+\frac{1}{8G_N}\left[3\log\left(\frac{r_b^2}{r_+^2}\right)+ (2\pi+|A|)r_+ \right]&(0<|A|<\frac{2\pi}{3}),\vspace{3mm}\\
\frac{3}{4G_N}\log\left[\frac{2}{\sqrt{3}}\right]+\frac{1}{8G_N}\left[3\log\left(\frac{r_b^2}{r_+^2}\right)+(4\pi-2|A|)r_+\right]&(\frac{2\pi}{3}<|A|<2\pi).
\end{cases}
\end{align}
Here the choice of $(n_1,n_2,n_3)$ would be $(n,n,n)$ for any integer $n$, which is equivalent to $(0,0,0)$ for $0<|A|<\frac{2\pi}{3}$ and $(0,-1,-1)$ for $\frac{2\pi}{3}<|A|<2\pi$. 

Second, the sum of entanglement entropy also has two different phases:
\begin{align}
\dfrac{1}{2}(S(A)+S(B)+S(C))&=\begin{cases}
\frac{1}{8G_N}\left[3\log\left(\frac{r_b^2}{r_+^2}\right)+2\pi r_+\right]&(0<|A|<\pi),\vspace{3mm}\\
\frac{1}{8G_N}\left[3\log\left(\frac{r_b^2}{r_+^2}\right)+(4\pi-2|A|)r_+\right]&(\pi<|A|<2\pi).
\end{cases}
\end{align}
Note that since $|B|=|C|<\pi$, we need to consider the phase transition only for $S(A)$. 
\begin{figure}[t]
    \centering
\includegraphics[width=15cm]{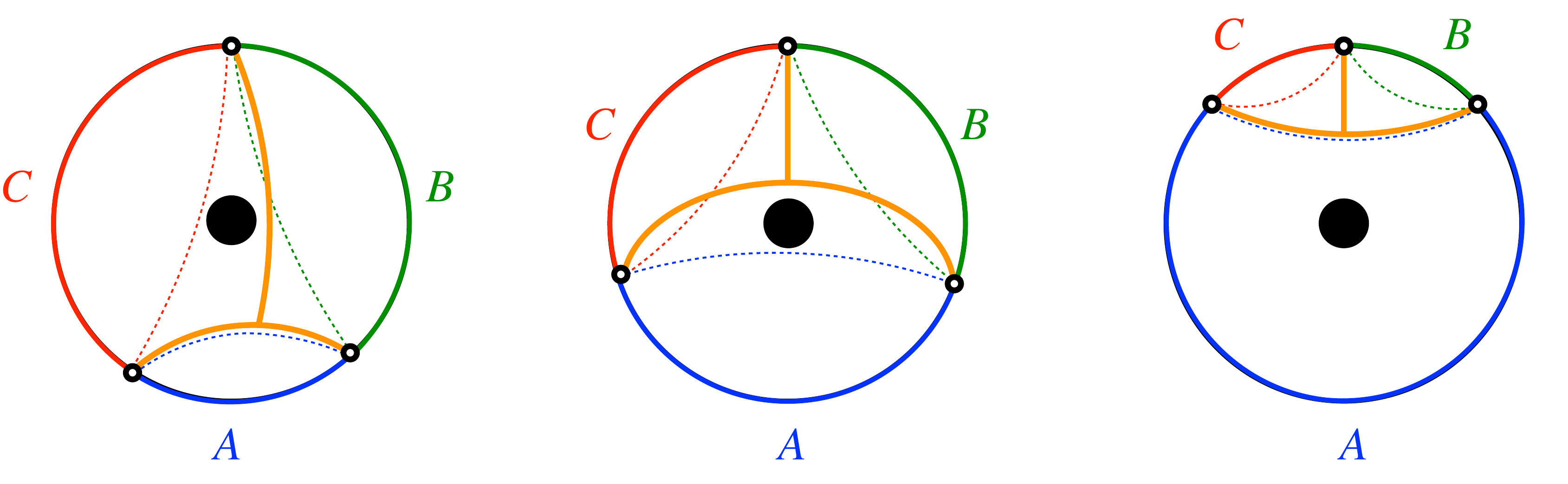}
    \caption{Three phases in the computation of ${\rm GM}^{(3)}$ around Eq.\eqref{eq:tri_eq}.
The left panel corresponds to $0<|A|<\frac{2\pi}{3}$. For presentation purposes, we slightly break the symmetry $|B|=|C|$ by taking $|B|<|C|$, so that the black hole lies in the domain $\mathcal{D}_{C}$.
The central panel corresponds to $\frac{2\pi}{3}<|A|<\pi$, where the black hole enters $\mathcal{D}_{A}$ as $A$ becomes the largest subsystem among the three.
The right panel corresponds to $\pi<|A|<2\pi$. In this regime, the RT surface computing $S(A)$ undergoes a phase transition. As a result, the computation of ${\rm GM}^{(3)}$ effectively reduces to that in global AdS${}_3$, and hence its leading contribution vanishes in the large-$r_+$ limit.}
    \label{fig:bh_tri}
\end{figure}
Therefore, ${\rm GM}^{(3)}$ has three phases as follows (see Figure \ref{fig:bh_tri}):
\begin{align}
{\rm GM}^{(3)}(A:B:C)&=\begin{cases}
\frac{3}{4G_N}\log\left[\frac{2}{\sqrt{3}}\right]+\frac{r_+|A|}{8G_N} &(0<|A|<\frac{2\pi}{3}),\\
\frac{3}{4G_N}\log\left[\frac{2}{\sqrt{3}}\right]+\frac{r_+(2\pi-2|A|)}{8G_N} &(\frac{2\pi}{3}<|A|<\pi),\\
\frac{3}{4G_N}\log\left[\frac{2}{\sqrt{3}}\right] &(\pi<|A|<2\pi).
\end{cases}\label{eq:tri_eq}
\end{align}
It is worth noting that once one of the three subsystems becomes larger than half of the total system, the tripartite entanglement ${\rm GM}^{(3)}$ vanishes up to the constant term common to vacuum AdS${}_3$. The maximum value is realized when $|A|=|B|=|C|$ as expected. The maximum value is given by 
\begin{align}
\label{maxGMtri}
\max\left[{\rm GM}^{(3)}(A:B:C)\right]=\frac{3}{4G_N}\log\left[\frac{2}{\sqrt{3}}\right]+\dfrac{1}{6}S_{BH},
\end{align}
where $S_{BH}$ is Bekenstein-Hawking entropy.
Assuming ${\rm GM}^{(3)}$ serves as a measure for tripartite entanglement, this implies that the static BTZ black hole has at most $\frac{1}{6}S_{BH}$ amount of tripartite entanglement. 

\subsection{tripartite (disconnected intervals)/fourpartite}\label{subsec:BTZtrid_four}
\begin{figure}[t]
    \centering
\includegraphics[width=7cm]{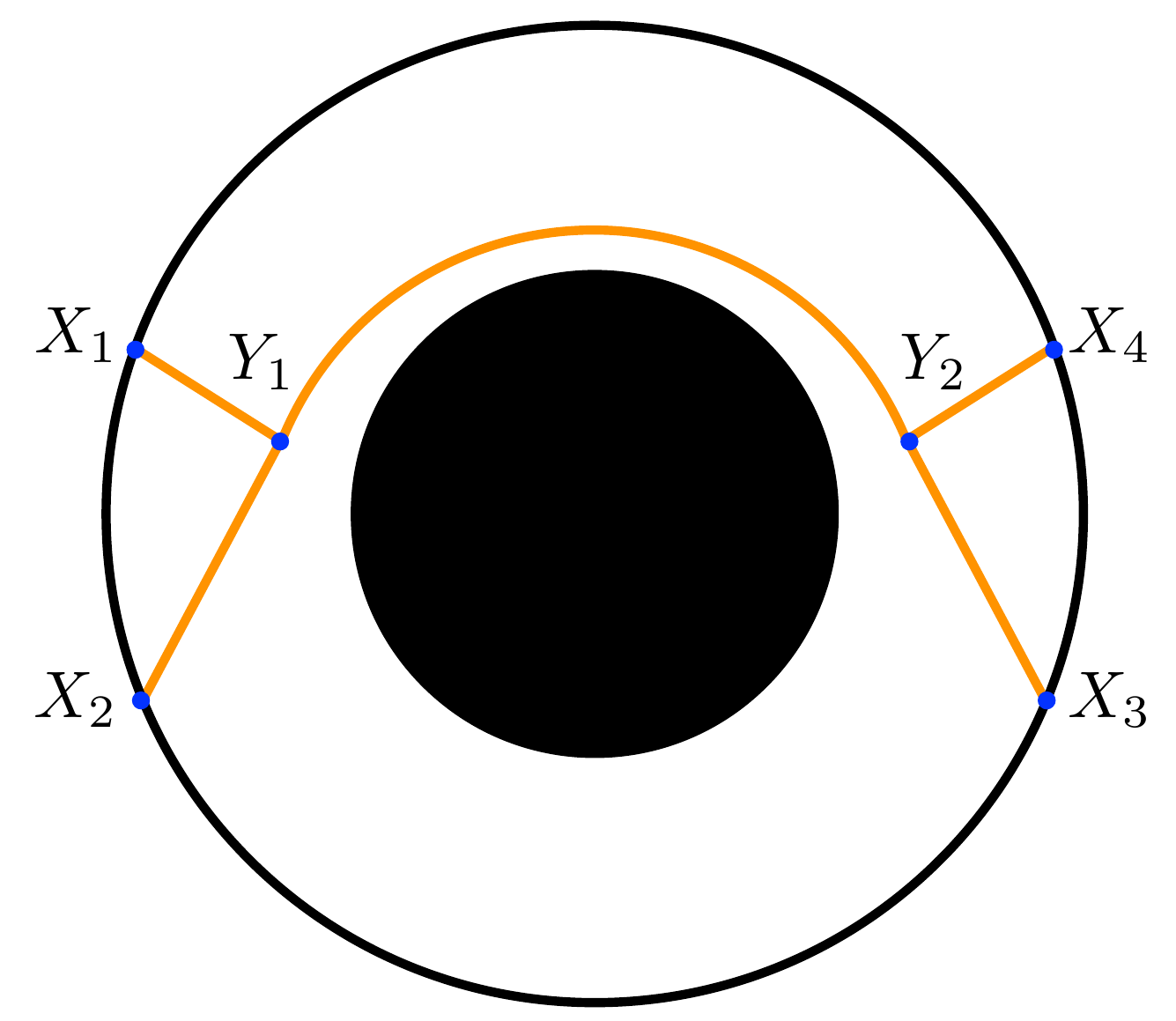}
    \caption{A tree-level Witten-like diagram relevant for computing $S^{(3)}$ with a disconnected interval or $S^{(4)}$ with adjacent intervals.
While we have possibly four phases for each channel depending on which of the four partitioned regions includes the black hole, we will focus on the symmetric setup where we do not need to care about this.}
    \label{fig:bh_four}
\end{figure}

Next, we calculate the length of $\mathcal{W}$ with two branch points like Figure \ref{fig:bh_four}. 
Let us take $\varphi_1 = |A|/2, \varphi_2 = -|A|/2, \varphi_3 = \pi-|A|/2, \varphi_4 = - \pi + |A|/2$. We assume that $0 < |A| < \pi$. And we denote the radius and angular position of $Y_1$ as $r$ and $\alpha$. Then, from the symmetry, $Y_2$ is located at radius $r$ and angular $\pi - \alpha$. For now, let us consider minimizing the length of $\mathcal{W}$ given below for $r$ and $\alpha$.
\begin{align}
\mathcal{W}(r,\alpha) = &\cosh^{-1}(-X_1^{(n_1)} \cdot Y_1^{(m_1)}) + \cosh^{-1}(-X_2^{(n_2)} \cdot Y_1^{(m_1)}) + \cosh^{-1}(-Y_1^{(m_1)}\cdot Y_2^{(m_2)}) \notag \\
& + \cosh^{-1}(-X_3^{(n_3)} \cdot Y_2^{(m_2)}) + \cosh^{-1}(-X_4^{(n_4)} \cdot Y_2^{(m_2)}). 
\end{align}
We may, by symmetry, restrict the parameter $\alpha$ to the range $-\frac{\pi}{2} < \alpha < \frac{\pi}{2}$. Depending on the sign of $\alpha$, the choice of integers $n_i$s that minimizes the quantity $\mathcal{W}(r,\alpha) $ differs.
For $-\frac{\pi}{2} < \alpha < 0$, the minimal value of $\mathcal{W}(r,\alpha) $ is obtained by choosing $n_1 = n_2 = n_3 - 1 = n_4 = m_1 = m_2$. On the other hand, for $0 < \alpha < \frac{\pi}{2}$, the minimal value of $\mathcal{W}(r,\alpha) $ is achieved by taking $n_1 = n_2 = n_3 = n_4 + 1 = m_1 = m_2 + 1$. Combining these cases, we find that, in the range $-\frac{\pi}{2} < \alpha < \frac{\pi}{2}$, the function to be minimized can be written as follows.
\begin{align}
\mathcal{W}(r,\alpha) &\approx 2\log(2r_b r_+^{-2}(r\cosh r_+\left(\frac{|A|}{2}-\alpha \right) - \sqrt{r^2-r_+^2})) \notag \\
&\ \ + 2\log(2r_b r_+^{-2}(r\cosh r_+ \left(\frac{|A|}{2}+\alpha \right) - \sqrt{r^2-r_+^2})) \notag \\
&\ \ \ \  + \cosh^{-1}(r_+^{-2}(r^2 \cosh r_+(\pi - 2 |\alpha|) - (r^2-r_+^2))) + O(r_b^{-2}).
\end{align}
Since this is an even function of $\alpha$, the values of $\alpha$ that minimize it always come in a pair with opposite signs.
These two choices correspond, respectively, to the trajectories that go around the black hole on the upper side and on the lower side after the branching.
Therefore, in what follows, we take $\alpha>0$ and focus on minimizing the function with respect to $\alpha$. Moreover, in what follows, for an analytic understanding, we assume large $r_b$ and, subsequently, also assume large $r_+$.

\begin{figure}[h]
 \centering
 \includegraphics[keepaspectratio, scale=0.7]{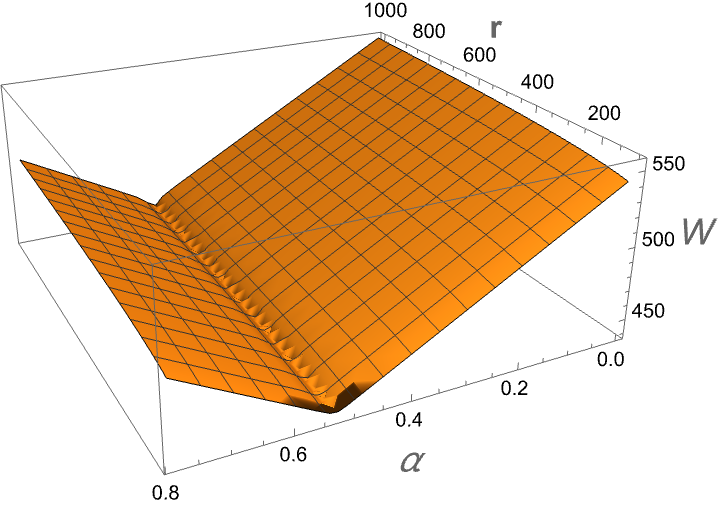}
 \caption{Length of geometry $\mathcal{W}(r,\alpha)$ as a function of $r$ and $\alpha$. Here we set $\frac{|A|}{2} = \pi/6 \approx 0.52$, $r_b = 10^4$ and $r_+ = 10^2$. Then $r_+/\ell$ is large enough, and it can be seen that the slope changes from one constant value to another with opposite sign around $\alpha \approx 0.52$.}
 \label{fig:Wplot3d}
\end{figure}

We first consider varying this function with respect to $\alpha$.
A direct differentiation with respect to $\alpha$ produces derivatives involving $\cosh$ and $\sinh$ functions.
Since these hyperbolic functions always contain $r_+$ in their arguments, they give rise to very large terms.
Keeping only such dominant contributions, the derivative can be approximated as follows.
\begin{align}
\frac{\partial \mathcal{W}(r,\alpha)}{\partial \alpha} &\approx -2r_+ \left\{\tanh\left(r_+\left(\pi-2\alpha\right)\right)+\tanh\left[r_+\left(\frac{|A|}{2}-\alpha \right)\right]-\tanh \left[r_+ \left(\frac{|A|}{2}+\alpha \right)\right] \right\} \\
&\approx
\begin{cases}
 -2r_+ & (\alpha = \frac{|A|}{2} + \epsilon),\\
 +2r_+ &  (\alpha = \frac{|A|}{2} - \epsilon),
\end{cases}
\end{align}
where $\epsilon$ is positive and starts from $r_+^0$ order (This means $r_+ \epsilon$ goes to infinity under the large $r_+$ limit). Therefore, at least in the large-$r_+$ limit, once $\alpha$ is separated from $|A|/2$ by more than an amount of order $1/r_+$, its slope becomes constant. See figure \ref{fig:Wplot3d}. From the above, we estimate $\alpha \approx |A|/2$ gives minimum value at large $r_+$ limit. In fact, when we set $\alpha = |A|/2$, we can verify $\frac{\partial \mathcal{W}(r,\alpha)}{\partial \alpha} \approx 0$ under the large $r_+$ limit.

\begin{figure}[h]
    \begin{minipage}[h]{0.45\hsize}
        \centering
        \includegraphics[width=.9\columnwidth]{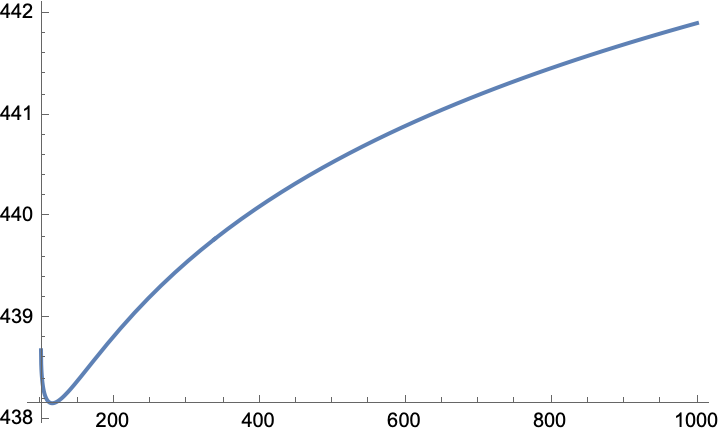}
    \end{minipage}
    \hspace{5mm}
    \begin{minipage}[h]{0.45\hsize}
        \centering
        \includegraphics[width=.95\columnwidth]{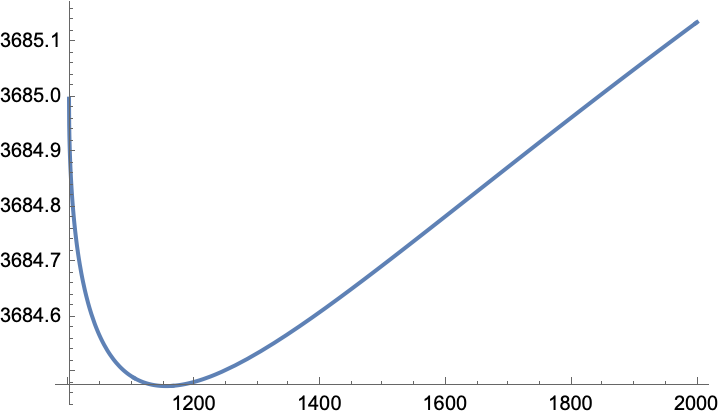}
    \end{minipage} \\
\caption{(left) Plot of $\mathcal{W}(r, \alpha = \frac{\pi}{6})$ when we set $r_b=10^4,\ r_+ = 10^2,\ \frac{|A|}{2} = \frac{\pi}{6}$. This becomes minimum at $r \approx 117$. 
(right) Plot of $\mathcal{W}(r, \frac{\pi}{6})$ when we set $r_b=10^5,\ r_+ = 10^3,\ \frac{|A|}{2} = \frac{\pi}{6}$. This becomes minimum at $r \approx 1153$.}
\label{fig:Wplotofr}
\end{figure}

Next, we consider minimizing with respect to $r$. Assuming that $r_+$ is large, we perform the variation and then substitute $\alpha = |A|/2$ to obtain the desired derivative.
This derivative also contains hyperbolic functions whose arguments involve variables proportional to $r_+$, and hence these contributions dominate.
Approximating each term in the limit where such dominant terms are much larger than all others, the result can be summarized as follows.
\begin{align}
\frac{\partial \mathcal{W}\left(r,|A|/2\right)}{\partial r} &\approx \frac{4}{r} + \frac{2}{r-\sqrt{r^2-r_+^2}} \left( 1 - \frac{r}{\sqrt{r^2-r_+^2}}\right).
\end{align}
This expression vanishes when $r = \frac{2}{\sqrt{3}}r_+ \approx 1.15 r_+$. This can also be verified through direct numerical calculation. See Figure~\ref{fig:Wplotofr}.

Therefore, at $r = \frac{2}{\sqrt{3}}r_+ $ and $\alpha = |A|/2$, $\mathcal{W}$ have the minimum value
\begin{align}
\mathcal{W}_{\rm min} \approx \log \left( \frac{4}{3} \right)^3  \frac{r_b^4}{r_+^4} + r_+(\pi+ |A|) + O(r_b^{-2},e^{-r_+}).
\end{align}
Here, the factor $e^{-r_+}$ in Landau symbol indicates that the subleading dependence on $r_+$ enters in an exponentially suppressed form, and the power in the exponent, namely the coefficient of $r_+$ appearing there, may be some constant.
From the above, $\mathcal{W}_{\rm min}$ has a term that grows linearly with $|A|$. This is because the geodesics from $X_2$ to the junction $Y_1$ near the horizon at angle $|A|/2$ contribute $r_+ |A|$ from the left and right, and the connecting geodesic contributes an additional length of $r_+(\pi - |A|)$.

\subsubsection*{As tripartite (disconnected intervals) system}
This $\mathcal{W}_{\min}$ admits two possible interpretations as a multi-entropy. The first is as a multi-entropy in the tripartite setup, in the case where one of the subsystems is disconnected. This corresponds to the phase shown in the right-hand panel of Figure~\ref{fig:S3bulkdis}: if the lengths of $A_1$ and $A_2$ are both $|A|$, then $\mathcal{W}_{\min}$ provides the bulk dual.
In this interpretation, once $|A|$ exceeds a certain critical value, $\mathcal{W}_{\min}$ splits into two RT surfaces (see the left panel of Figure~\ref{fig:S3bulkdis}). We refer to this phase as the RT-surface phase. The value of $S^{(3)}$ in the RT-surface phase is 
\begin{align}
4G_N S^{(3)} = 2\log  \frac{4r_b^2}{r_+^2} \left[ (\sinh^2 \frac{r_+}{2} (\pi - |A|) ) \right] \approx 2\log  \frac{r_b^2}{r_+^2}+ 2 r_+ (\pi - |A|) ),
\end{align}
and the transition occurs at some value
\begin{align}
|A|_{\rm cr} \approx \frac{\pi}{3} - \frac{1}{r_+}\log \frac{4}{3} + O(r_b^{-2},e^{-r_+}).
\end{align}
The genuine multi-entropy can be evaluated as 
\begin{align}
{\rm GM}^{(3)}_{\rm BTZ}(A_1A_2:B:C) = \frac{1}{4G_N} \times
\begin{cases}
3 \log \frac{4}{3} + r_+ |A| & 0<|A|\leq |A|_{\rm cr},\\
0 & |A|_{\rm cr} \leq |A| < \pi.\\
\end{cases}
\end{align}
The maximum value is realized when $|A|_{\rm cr}$. Up to the leading order of $r_+$, the maximum value is given by
\begin{align}
{\rm max}[{\rm GM}^{(3)}_{\rm BTZ}(A_1A_2:B:C)] = \frac{4}{4G_N} \log \left[\frac{2}{\sqrt{3}} \right]+ \frac{1}{6}S_{BH}.
\end{align}
First, it is important to note that, as in equation (\ref{maxGMtri}), the size-dependent term is $\frac{1}{6} S_{BH}$. Therefore, even in disconnected cases, tripartite entanglement is at most $\frac{1}{6} S_{BH}$, excluding the universal constant term.
And figure~\ref{fig:MaxGM3} shows the results of a numerical analysis for finite values of $r_+$.
When $r_+$ is relatively large, both $\max [\text{GM}^{(3)}]$ and the critical value $|A|_{\mathrm{cr}}$ are seen to asymptotically approach the analytic expressions obtained above.

Let us comment on the coefficient of the constant term. Note that, unlike in (\ref{maxGMtri}), the coefficient is different here. In the tripartite case where the system is divided into three connected regions, this constant term was equal to $3 \times \frac{1}{4G_N} \log \frac{2}{\sqrt{3}}$.
By contrast, in the present setup, where the system is divided into three parties with one of them being disconnected, the constant term becomes $4 \times \log \frac{1}{4G_N} \frac{2}{\sqrt{3}}$.
From this observation, one is led to conjecture that the constant term takes the following form:
\begin{align}
\text{the constant term in }&{\rm max}[{\rm GM}^{(3)}_{\rm BTZ}(A:B:C)] \notag \\
&= (\text{\# of anchoring point in boundary}) \times \frac{1}{4G_N} \log \left[ \frac{2}{\sqrt{3}}\right].
\end{align}
It is suggestive that such a term follows an ``area law'' as in the bipartite case. A detailed analysis of this point is left for future work.

\subsubsection*{Comparison with AdS${}_3$}
For comparison, let us introduce the multi-entropy for the same setup (tripartite setup with disconnected interval) in the global $\mathrm{AdS}_3$.
When the sizes of $A_1$ and $A_2$ are small, the corresponding $\mathcal{W}_{\rm min}$ (whose shape is the one in (\ref{fig:S4dual})) has been computed in \cite{Gadde:2023zzj} as follows
\begin{align}
\mathcal{W}_{\rm min} =  \log 2 r_b^4 + 6 \log \frac{2}{\sqrt{3}} +  \log  (1-\cos |A|) \left(1+\cos \frac{|A|}{2}\right)^2\ \ \ \ \text{for AdS$_3$}
\end{align}
The value of $S^{(3)}$ in the RT-surface phase is 
\begin{align}
4G_N S^{(3)} = 2 \log 2r_b^2(1+\cos |A|),
\end{align}
and the transition occurs at some value which satisfies
\begin{align}
\cos \frac{|A|_{\rm cr}}{2} \approx 0.858,\ \ \ |A|_{\rm cr} \approx 1.08\ (\simeq 0.344 \pi).
\end{align}
The genuine multi-entropy can be evaluated as 
\begin{align}
{\rm GM}^{(3)}_{\rm AdS}(A_1A_2:B:C) = \frac{1}{4G_N} \times
\begin{cases}
3 \log \frac{4}{3} + 2\log \frac{1+\cos \frac{|A|}{2}}{\sqrt{2(1+\cos |A|)}} & 0<|A|\leq |A|_{\rm cr}\\
0 & |A|_{\rm cr} \leq |A| < \pi\\
\end{cases}
\end{align}
Then the maximum is given by
\begin{align}
{\rm max}[{\rm GM}^{(3)}_{\rm AdS}(A_1A_2:B:C)] = \frac{4}{4G_N} \log \left[\frac{2}{\sqrt{3}} \right]+ \frac{1}{4G_N}\times 0.447 \cdots .
\end{align}
\begin{figure}[h]
    \begin{minipage}[h]{0.45\hsize}
        \centering
        \includegraphics[width=1\columnwidth]{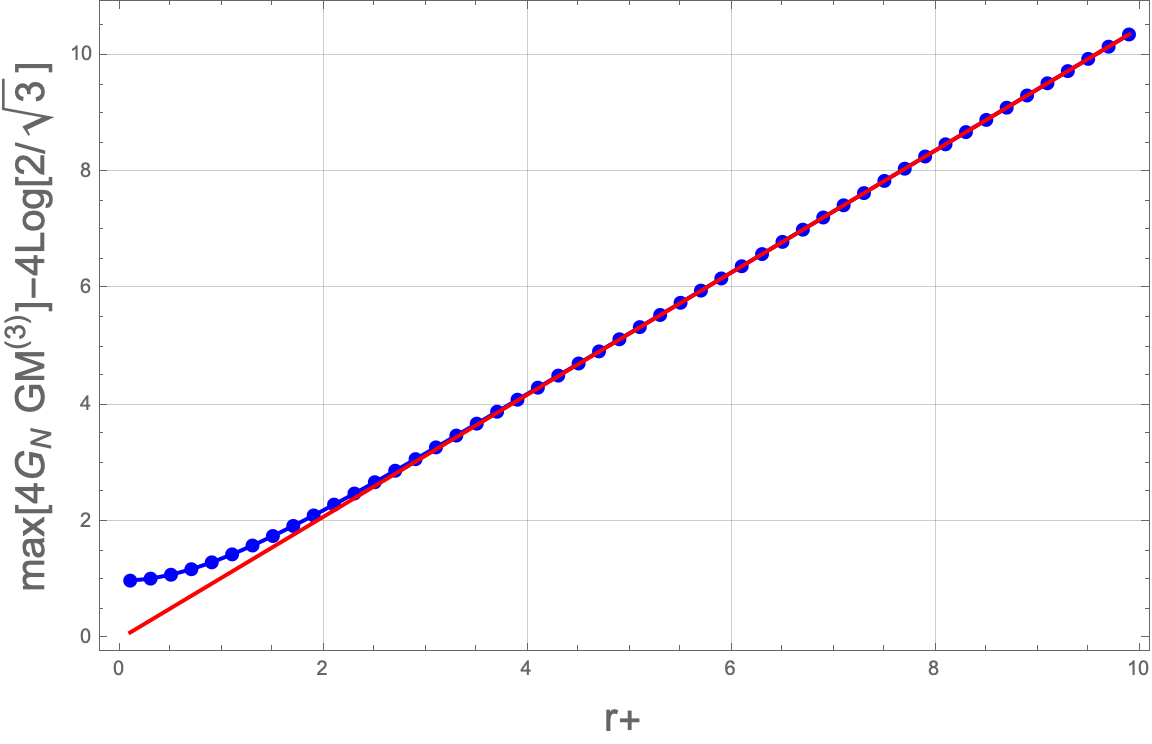}
    \end{minipage}
    \hspace{5mm}
    \begin{minipage}[h]{0.45\hsize}
        \centering
        \includegraphics[width=1\columnwidth]{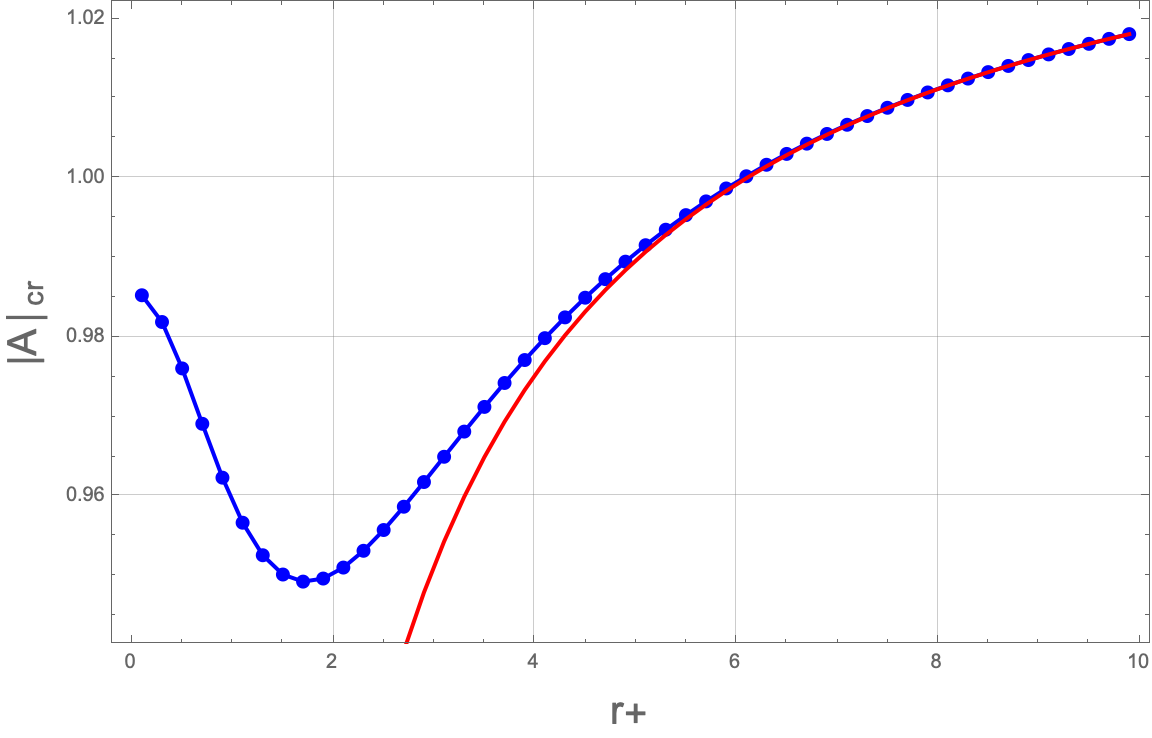}
    \end{minipage} \\
\caption{(left) Plot of max$[4G_N \text{GM}_{\rm BTZ}^{(3)}(A_1A_2:B:C)] - 4 \log \left[\frac{2}{\sqrt{3}}\right]$ as a function of $r_+$. Red line represents $4G_N \times\frac{1}{6}S_{BH} = \frac{\pi}{3}r_+$.  
(right) Plot of $|A|_{\rm cr}$ as a function of $r_+$. Red line represents $\frac{\pi}{3} - \frac{1}{r_+}\log \frac{4}{3}$.}
\label{fig:MaxGM3}
\end{figure}
To compare with any size of BTZ, it is necessary to rely not only on the large-$r_+$ asymptotic forms discussed above, but also on numerical calculations at finite $r_+$.
The results are shown in Figure~\ref{fig:MaxGM3}. From the left figure, one observes that for small $r_+$ the behavior differs from that in the large-$r_+$ regime. This is because we have set $\ell = 1$, and the qualitative behavior changes around $r_+ \sim 1$.
As a consequence, this quantity does not approach zero as $r_+ \to 0$, but instead remains finite.
Although the $r_+ \to 0$ limit of the BTZ geometry does not directly reduce to global AdS spacetime, the persistence of a finite value can be regarded as indirect evidence supported by this numerical analysis. 

From the right figure, one also finds that the critical value $|A|_{\mathrm{cr}}$, which decreases as $r_+$ becomes smaller, eventually turns around and starts to increase as $r_+ \to 0$. Physically, this can be understood as follows.
As $r_+$ decreases, the surface $\mathcal{W}_{\rm min}$ can pass closer to the center of the BTZ spacetime, which tends to reduce its length. As a result, a larger value of $|A|$ is required for the configuration to transition to the RT-surface phase. As a result, it is reasonably understandable that in $\mathrm{AdS}_3$ the critical value $|A|_{\mathrm{cr}}$ exceeds $\frac{\pi}{3}$.

\subsubsection*{As fourpartite system}
As a second interpretation of the length of $\mathcal{W}$, one may regard this as the multi-entropy in the fourpartite case.
Here the system is divided into $A,B,C,$ and $D$ as shown in (\ref{fig:S4dual}), and we are computing the bulk dual of the multi-entropy in a symmetric configuration where $A$ and $B$ have equal sizes, $|A|=|B|$, and likewise $C$ and $D$ are equal in size.

In this case, as $|A|$ increases, the configuration does not transition to a simple RT surface.
Instead, it undergoes a transition from a $t$-channel–type $\mathcal{W}$ to an $s$-channel–type $\mathcal{W}$.
This transition occurs at $|A|=\pi/2$. Therefore,
\begin{align}
S^{(4)}(A:B:C:D) = \frac{1}{4G_N}
\begin{cases}
\log \left( \frac{4}{3} \right)^3  \frac{r_b^4}{r_+^4} + r_+(\pi+ |A|) & 0<|A|<\frac{\pi}{2}\\
\log \left( \frac{4}{3} \right)^3  \frac{r_b^4}{r_+^4} + r_+(2\pi- |A|) & \frac{\pi}{2}<|A|<\pi
\end{cases}
\end{align}
First, we study the entropy excluded by only bipartite entanglement of a single subsystem (This is discussed in \cite{Iizuka:2025ioc}.)
\begin{align}
{\rm GM}^{(3,4)} &\equiv S^{(4)} - \frac{1}{2}(S(A)+S(B)+S(C)+S(D)).
\end{align}
Since the size of each region is smaller than $\pi$, the standard formula for the bipartite entanglement entropy, $S(A) = \dfrac{1}{4G_N}\log\left[\frac{4r_b^2}{r_+^2}\sinh^2\left(\frac{r_+}{2}\varphi_{ij}\right)\right]$ can be used without modification.
Therefore, we evaluate the ${\rm GM}^{(3,4)}$ as follows.
\begin{align}
4G_N {\rm GM}^{(3,4)} & =  \log \left( \frac{4}{3} \right)^3  \frac{r_b^4}{r_+^4} + r_+(\pi+ |A|) \notag \\
&\ \ \ \ \ \  -\log\left[\frac{4r_b^2}{r_+^2}\sinh^2 \frac{r_+}{2} |A|\right] - \log\left[\frac{4r_b^2}{r_+^2}\sinh^2\left(\frac{r_+}{2}(\pi-|A|)\right)\right] \notag \\
& \approx 3 \log \frac{4}{3} + r_+ |A|
\end{align}
This is justified in the range of $r_b \gg r_+ \gg \ell = 1$ and $|A|$ is large enough such that $r_+ |A| \gg 1$ holds. In addition, ${\rm GM}^{(3,4)}$ contains contributions from both tripartite and fourpartite entanglement. By subtracting the genuine tripartite contribution, one can isolate the genuine fourpartite entanglment.

\begin{figure}[htbp]
\begin{minipage}[t]{0.5\hsize}
        \centering
    \includegraphics[width=\columnwidth]{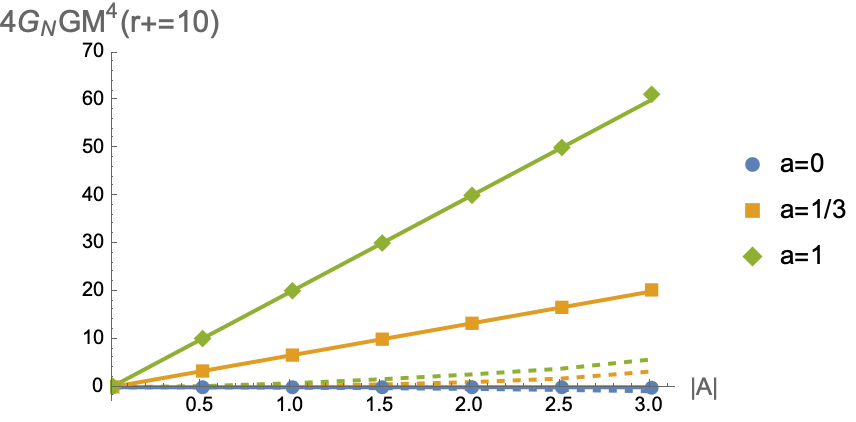}
      \end{minipage}
      \hspace{5mm}
      \begin{minipage}[t]{0.5\hsize}
        \centering
    \includegraphics[width=\columnwidth]{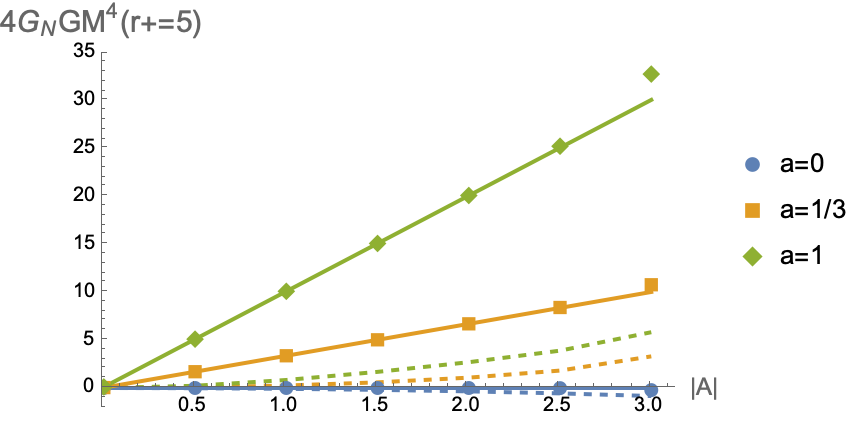}
      \end{minipage} \\
\begin{minipage}[t]{0.5\hsize}
        \centering
    \includegraphics[width=\columnwidth]{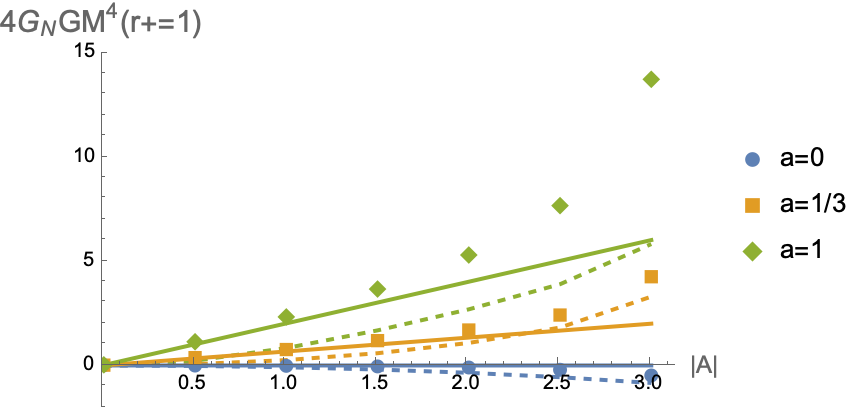}
      \end{minipage}
      \hspace{5mm}
      \begin{minipage}[t]{0.5\hsize}
        \centering
    \includegraphics[width=\columnwidth]{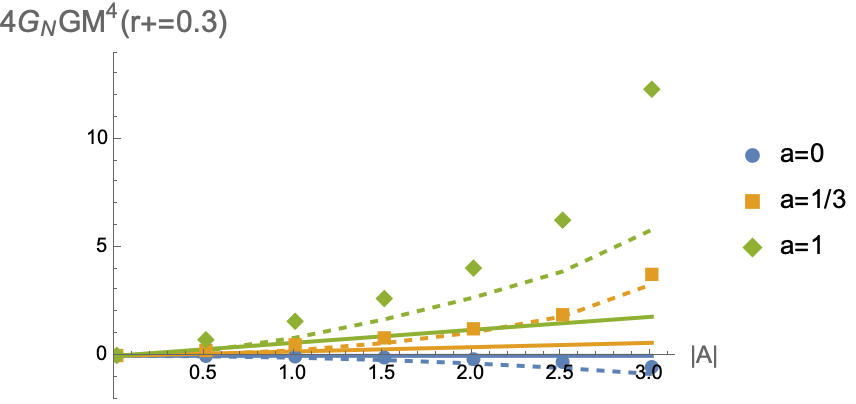}
      \end{minipage} \\
      \caption{Plot of $\text{GM}^{(4)}$ in a particular phase ($t$-channel) as a function of $|A| \in (0,\pi)$ in the BTZ background for $r_+=10,\,5,\,1,\,0.3$. For each plot, the cases $a=0,\,1/3,\,1$ are shown in different colors. The solid straight lines represent the large-$r_+$ limiting form (\ref{GM4forBTZ}), while the dots of the same color show the corresponding numerical results at finite $r_+$. The dashed curves show $\text{GM}^{(4)}$ in global $\mathrm{AdS}_3$, using the same color scheme for $a=0,\,1/3,\,1$. Furthermore, assuming a large $r_b$ limit, GM${}^{(4)}$ is analytically independent of $r_b$ even for finite $r_+$. Therefore, $r_b$ was not specified in this numerical calculation (effectively taken as very large).}
      \label{PlotofGM4}
\end{figure}

As with the tripartite entanglement among $AB$, $C$ and $D$, we should subtract all ${4 \choose 2}=6$ tripartite terms. One of them is the case where the tripartite contribution coincides with the fourpartite one, in which case ${\rm GM}^{(3)}$ is given by ${\rm GM}^{(3)}(AB:C:D) = S^{(4)} - \frac{1}{2}(S(AB) + S(C) + S(D))$. Another case is where the tripartite contribution is completely separated into geodesics and represented by two RT surfaces; in that case, ${\rm GM}^{(3)}(CD:A:B) = (S(A)+S(B)) - \frac{1}{2}(S(CD) + S(A) + S(B)) = 0$. The remaining genuine tripartite can be evaluated as follows. For example, consider $S^{(3)}(BC:A:D)$. The size of $BC$ is $\pi$, the size of $A$ is $|A|$, which ranges from $0$ to $\pi$, and finally the size of $D$ is $\pi-|A|$, which also lies between $0$ and $\pi$. Therefore, none of the subtleties discussed in Section~\ref{sec:BTZtri} arise in this case. More explicitly, we are in the situation where we have to minimize (\ref{eq:tri-minimize}) and in this regime the choice $n_1 = n_2 = n_3$ minimizes each contribution and hence also minimizes their sum. As a result, the second term on the right-hand side of (\ref{eq:tri-minimize}) can be interpreted simply as the sum of RT surfaces, $\frac{1}{2}\bigl(S(BC)+S(A)+S(D)\bigr)$, and we obtain
\begin{align}
{\rm GM}^{(3)}(BC:A:D) = \frac{3}{4G_N}\log\frac{2}{\sqrt{3}}.
\end{align}
The situations for ${\rm GM}^{(3)}(AC:B:D)$, ${\rm GM}^{(3)}(AD:B:C)$, and ${\rm GM}^{(3)}(BD:A:C)$ are completely identical, and hence all of them yield the same value. From the above, we can calculate the genuine fourpartite multi-entropy as follows.
\begin{align}
&{\rm GM}^{(4)}_{\rm BTZ}(A:B:C:D)  = \frac{2}{3}S^{(4)}(A:B:C:D) - \frac{4}{3} \times {\rm GM}^{(3)}(BC:A:D) \notag \\
&\ \ + \left(a-\frac{1}{3}\right)(S(AB)+S(AC)+S(AD)) - a(S(A)+S(B)+S(C)+S(D))\\
\label{GM4forBTZ}
&= \frac{ar_+}{2G_N}|A|
\end{align}
From this result, the maximum value of ${\rm GM}^{(4)}_{\rm BTZ}(A:B:C:D)$ is realized at $|A| = \frac{\pi}{2}$ and the value is
\begin{align}
\text{max}\left[{\rm GM}^{(4)}_{\rm BTZ}(A:B:C:D)\right]= \frac{a}{2}S_{BH}.
\end{align}
This is the result in large $r_+$ limit. Figure~\ref{PlotofGM4} presents the explicit numerical results, including the cases with finite $r_+$. Here we assume the $t$-channel configuration for $\mathcal{W}$, and, using its length, we plot the resulting $\text{GM}^{(4)}$ over the range $0 \le |A| \le \pi$ for completeness.
In practice, however, $\mathcal{W}$ undergoes a transition to s-channel at $|A|=\pi/2$.
Therefore, the actual behavior of $\text{GM}^{(4)}$ for $\pi/2 \le |A| \le \pi$ is that the behavior on the interval $0 \le |A| \le \pi/2$ is reflected about $|A|=\pi/2$.

Computing the same quantity in the case of global $\mathrm{AdS}_3$, one finds
\begin{align}
{\rm GM}^{(4)}_{\rm AdS}(A:B:C:D)  = \frac{4}{3} \left( 2 \log \left[\cos \frac{|A|}{4}\right] - 3a \log \left[\cos \frac{|A|}{2} \right]\right).
\end{align}
See Figure~\ref{PlotofGM4} again to compare these results.
The important point is that, for fixed $a$, the global $\mathrm{AdS}_3$ result (dashed curves) does not exceed the corresponding data dots for finite $r_+$ BTZ case in all plots. This means that the BTZ black hole tends to have more four-partite entanglement than global AdS${}_3$.

\subsubsection*{Comment on the free parameter in ${\rm GM}^{(4)}$}
We conclude this section by commenting on the parameter $a$, which remains undetermined in the definition of $\text{GM}^{(4)}$. We find that as $a$ increases, $\text{GM}^{(4)}$ tends to increase in both the global AdS${}_3$ and BTZ cases in numerical calculations.
This can be understood from (\ref{GM4def}), since in the present setup where $ABCD$ forms a pure state one can write 
\begin{align}
\frac{\partial}{\partial a} {\rm GM}^{(4)}(A:B:C:D) = - I_3(A:B:C),
\end{align}
where $I_3(A:B:C)$ is the tripartite mutual information.
For states that admit a holographic dual, the right-hand side is shown to be non-negative~\cite{Hayden:2011ag}, while it can be positive for more general states. We also note that, depending on the value of $a$, $\text{GM}^{(4)}$ itself can become negative. A more careful investigation of the parameter $a$ is expected to lead to an unambiguous definition of $\text{GM}^{(4)}$. We leave this issue for future work.

\section{Finite $r_b$ effect}\label{sec:finitecutoff}
We know that ${\rm GM}^{(3)}$ for adjacent intervals in vacuum AdS${}_3$ (ground state in CFT${}_2$) remains constant regardless of the subsystem size. This raises a naive question of whether holographic multi-entropy and its variant capture tripartite entanglement quantitatively. 
In this section, by introducing a finite cutoff $r_b$, we demonstrate that a nontrivial size-dependence emerges: as we decrease $r_b$, the value of ${\rm GM}^{(3)}$ monotonically decreases. Such a finite-cutoff holography can be realized in the context of AdS/CFT correspondence with $T\overline{T}$-deformation~\cite{McGough:2016lol}. The constant piece itself may be interpreted as a universal multipartite entanglement originating from the conformal invariance of the UV (CFT). However, deeper in the IR, the system appears to encode multipartite entanglement across various scales.
We also perform a parallel analysis for the BTZ black hole and identify how the tripartite entanglement is modified in that geometry.

We first analyze the perturbative behavior in an expansion in $r_b$, and subsequently examine the full-order behavior through numerical plots.
Expanding the boundary coordinates $X_i$ in powers of $r_b$, we obtain
\begin{align}
X_{iA} = r_b P_{iA}+\frac{1}{r_b}Q_{iA}+\mathcal{O}(r^{-3}_b).
\end{align}
This leads to the following expansion for the inner products between boundary points:
\begin{align}
X_{ij}\equiv -X_i\cdot X_j=r_b^2 P_{ij}+r_b^0 N_{ij}+\mathcal{O}(r_b^{-2}),
\end{align}
Since we will focus on an equal-time slice, $t_i=t_j$, the sub-leading term simplifies to
\begin{align}
N_{ij}=-(P_i\cdot Q_j+P_j\cdot Q_i)=1.
\end{align}
This simplification applies to both the AdS${}_3$ and BTZ backgrounds. For later convenience, however, we keep $N_{ij}$ arbitrary for the moment.
Next, we need the inner products between $Y_i$ and each $X_j$. For this purpose, we expand
\begin{align}
y_i\equiv -Y\cdot X_i=r_b(-Y\cdot P_i)+r_b^{-1}(-Y\cdot Q_i)+\mathcal{O}(r^{-3}_b)\equiv r_b p_i+r_b^{-1} q_i+\mathcal{O}(r_b^{-3}) .
\end{align}
Plugging these expressions into \eqref{eq:minimal}, we obtain the leading-order equations
\begin{align}
3p_j-\sum_{i=1}^3 \frac{P_{ij}}{p_i}=0 ,
\end{align}
which arise at $\mathcal{O}(r_b)$, and the next-to-leading-order equations
\begin{align}
3q_j + \sum_{i=1}^3 \dfrac{1}{2p_i^3} \left[p_i(p_j-2N_{ij}p_i)+P_{ij}(-1+2p_i q_i)\right] = 0 ,
\end{align}
which appear at $\mathcal{O}(r_b^{-1})$.
We then solve these equations order by order. The $\mathcal{O}(r_b)$ equations have already been solved previously, but for usefulness, we summarize the solutions here: 
\begin{align} 
p_1=\sqrt{\dfrac{2}{3}\dfrac{P_{12}P_{13}}{P_{23}}},\quad p_2=\sqrt{\dfrac{2}{3}\dfrac{P_{12}P_{23}}{P_{13}}},\quad p_3=\sqrt{\dfrac{2}{3}\dfrac{P_{13}P_{23}}{P_{12}}}.
\end{align}
Then, we obtain the sub-leading order solutions as
\begin{align}
q_1 &= \frac{P_{23}^2 - 2P_{12}P_{13}N_{23} + 2P_{23}(P_{12}N_{13}+P_{13}N_{12})}{2\sqrt{6} \sqrt{P_{12}P_{13}P_{23}^3}}\\
q_2 &= \frac{P_{13}^2 - 2P_{12}P_{23}N_{13} + 2P_{13}(P_{12}N_{23}+P_{23}N_{12})}{2\sqrt{6} \sqrt{P_{12}P_{23}P_{13}^3}}\\
q_3 &= \frac{P_{12}^2 - 2P_{13}P_{23}N_{12} + 2P_{12}(P_{13}N_{23}+P_{23}N_{13})}{2\sqrt{6} \sqrt{P_{13}P_{23}P_{12}^3}}
\end{align}
Assuming $N_{ij}=1$ as mentioned above, we obtain the large $r_b$ expansion of $S^{(3)}(A:B:C)$ as
\begin{align}
S^{(3)}(A:B:C) &= \dfrac{1}{4G_N}\sum_{i=1}^3 \left[\log(2p_i r_b^2)+\dfrac{4p_iq_i-1}{4p_i^2 r_b^2}\right]+\mathcal{O}(r_b^{-4})
\end{align}
In particular, $\mathcal{O}(r_b^{-2})$ part is expressed as
\begin{align}
-\dfrac{1}{4G_N}\dfrac{(P_{12}^2+P_{23}^2+P_{13}^2)-4(P_{12}P_{23}+P_{23}P_{13}+P_{13}P_{12})}{8 P_{12}P_{13}P_{23}}\dfrac{1}{r_b^2}
\end{align}
In what follows, we discuss AdS${}_3$ and the static BTZ case in turn. 
\subsection{AdS${}_3$ case}
\begin{figure}[tbp]
    \centering
\includegraphics[width=9cm]{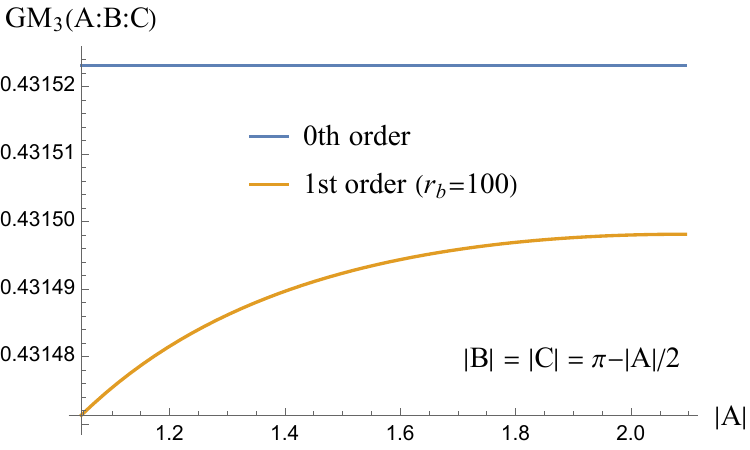}
    \caption{The maximum value is realized when $|A|=|B|=|C|$. Plots for small $|A|$ are not shown, where the approximation is no longer reliable.}
    \label{fig:plot_large-r_AdS}
\end{figure}

\begin{figure}[tbp]
    \centering
\includegraphics[width=9cm]{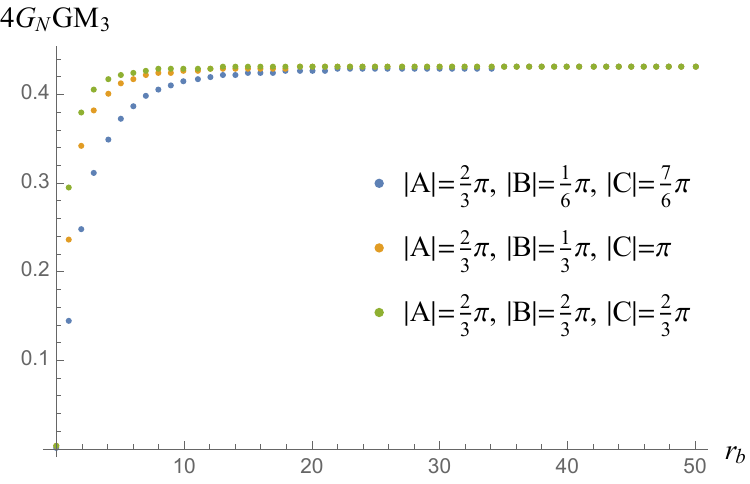}
    \caption{Plots for ${\rm GM}^{(3)}$ at finite-$r_b$ in AdS${}_3$. The maximum value is always realized when $|A|=|B|=|C|$. }
    \label{fig:finiter_AdS}
\centering\vspace{1cm}
\includegraphics[width=12cm]{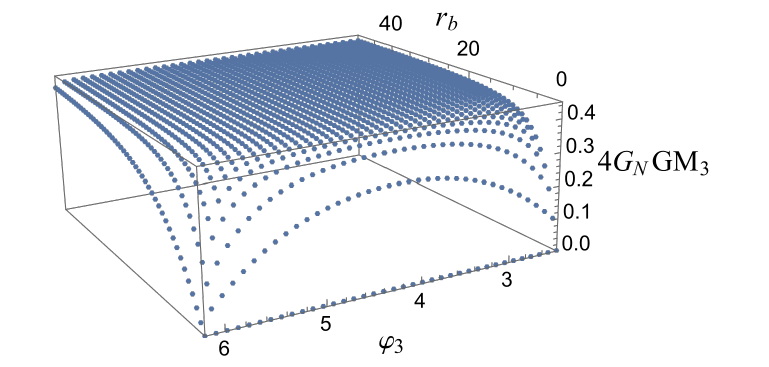}
    \caption{Full $r_b$ and $\varphi_3$-dependence of ${\rm GM}^{(3)}$ for AdS${}_3$. Here we fixed $|A|=\frac{2}{3}\pi$, $|B|=\varphi_3-\frac{2}{3}\pi>0$ and $|C|=2\pi-\varphi_3>0$.}
\label{fig:finiter_AdS_full}
\end{figure}

We first discuss the AdS${}_3$ case. Since the candidate of the minimal surface is unique in this case, we obtain\footnote{One may be interested in whether the same result can be reproduced from CFT${}_2$ side using perturbative expansion of the coupling in $T\overline{T}$-deformation. However, we cannot see a perfect match in this simple way as we need to improve OPE for stress tensor and twist operators~\cite{Lewkowycz:2019xse}. We thank Mitsuhiro Nishida for discussing this point.}
\begin{align}
{\rm GM}^{(3)}(A:B:C)=\frac{3}{4G_N}\log\left[\frac{2}{\sqrt{3}}\right] - \frac{1}{4G_N} \frac{P_{12}^2 + P_{13}^2 + P_{23}^2}{8(P_{12}P_{13}P_{23})} \frac{1}{r_b^2}+\mathcal{O}(r_b^{-4}),
\end{align}
where
\begin{align}
P_{ij}&=
2\sin^2\frac{\varphi_{i}-\varphi_{j}}{2}.
\end{align}
In particular, the sub-leading part depends on the subsystem sizes and makes ${\rm GM}^{(3)}$ smaller. The maximum ${\rm GM}^{(3)}$ up to $\mathcal{O}(r_b^{-2})$ is realized when three subsystems have the same size, \textit{i.e.}, $|\varphi_i-\varphi_j|=\frac{2\pi}{3}$ for all $1\leq i<j\leq 3$. See Figure \ref{fig:plot_large-r_AdS}. Indeed, this suggests a version of ``mutuality'' for tripartite entanglement and ${\rm GM}^{(3)}$ indeed measures it. As is obvious from the geometric picture, ${\rm GM}^{(3)}$ reaches to zero as $r_b\rightarrow0$ limit. 
For small $r_b$, ${\rm GM}^{(3)}$ grows linearly in $r_b$. We leave the intermediate calculation in appendix \ref{app:smallr}. 
One can also solve \eqref{eq:minimal} numerically and see full $r_b$-dependence. See Figure \ref{fig:finiter_AdS} and \ref{fig:finiter_AdS_full}. 

\subsection{BTZ case}
Keeping in mind the possibility of the phase transition discussed in Section~\ref{sec:BTZ}, we can also investigate the finite-$r_b$ behavior of the holographic multi-entropy in the pure BTZ black hole. 
For the large $r_+$ limit, however, these contribution is exponentially suppressed, so we will be a bit brief about the analytic expression. 

The $\mathcal{O}(r_b^{-2})$ part of $S^{(3)}(A:B:C)$ can be expressed as
\begin{align}
\delta S^{(3)}_{r_b^{-2}}(\{\varphi_i\},\{n_j\})&=-\dfrac{1}{4G_N}\dfrac{(P_{12}^2(n_{12})+P_{23}^2(n_{23})+P_{13}^2(n_{13}))}{8P_{12}(n_{12})P_{13}(n_{13})P_{23}(n_{23})}\dfrac{1}{r_b^2}\nonumber\\
&+\dfrac{1}{4G_N}\dfrac{P_{12}(n_{12})P_{23}(n_{23})+P_{23}(n_{23})P_{13}(n_{13})+P_{13}(n_{13})P_{12}(n_{12})}{2 P_{12}(n_{12})P_{13}(n_{13})P_{23}(n_{23})}\dfrac{1}{r_b^2}, \label{eq:sub_btz}
\end{align}
where
\begin{align}
P_{ij}(n_{ij})=2\sinh^2\dfrac{r_+}{2}(\varphi_i-\varphi_j+2\pi n_{ij}). 
\end{align}
Here, each $n_{ij}=n_i-n_j$ is already fixed by the minimization of the leading order contribution discussed in Section~\ref{sec:BTZ}. 

For AdS$_3$, the second line of Eq.~\eqref{eq:sub_btz} cancels out completely when evaluating
${\rm GM}^{(3)}(A\!:\!B\!:\!C)$.
For pure BTZ, however, since the choice of $n_{ij}$ for $S^{(3)}(A\!:\!B\!:\!C)$ differs from that
for $S(A)$, $S(B)$, and $S(C)$, this cancellation does not occur.

Following the example in the previous section, let us again vary the size of $|A|$ and assume $|B|=|C|(=\pi-\frac{1}{2}|A|)$. In this case, we have three phases:
\begin{align}
&\delta{\rm GM}^{(3)}(A:B:C)|_{r_b^{-2}}\nonumber\\
&=
\begin{cases}
\delta S^{(3)}_{r_b^{-2}}(\{\varphi_i\},\{n_j\})|_{n_1=n_2=n_3=0}-\overline{\delta S^{(2)}_{r_b^{-2}}(\{\varphi_i\},\{m_{ij}\})}|_{m_{12}=m_{23}=m_{31}=0}&(0<|A|<\frac{2\pi}{3})\\
\delta S^{(3)}_{r_b^{-2}}(\{\varphi_i\},\{n_j\})|_{n_1=0, n_2=n_3=-1}-\overline{\delta S^{(2)}_{r_b^{-2}}(\{\varphi_i\},\{m_{ij}\})}|_{m_{12}=m_{23}=m_{31}=0}& (\frac{2\pi}{3}<|A|<\pi)\\
\delta S^{(3)}_{r_b^{-2}}(\{\varphi_i\},\{n_j\})|_{n_1=0, n_2=n_3=-1}-\overline{\delta S^{(2)}_{r_b^{-2}}(\{\varphi_i\},\{m_{ij}\})}|_{m_{12}=1, m_{23}=m_{31}=0} & (\pi<|A|<2\pi)
\end{cases}
\end{align}
where
\begin{align}
\overline{\delta S^{(2)}_{r_b^{-2}}(\{\varphi_i\},\{m_{ij}\})}&\equiv\frac{1}{2}(S(A)+S(B)+S(C))=\dfrac{1}{4G_N}\dfrac{1}{2}\sum_{i<j}\dfrac{1}{P_{ij}(m_{ij})}
\end{align}
and $m_{ij}$ is already fixed from the leading order contribution of $S(A), S(B)$ and $S(C)$. 
\begin{figure}[htbp]
    \begin{minipage}[h]{0.45\hsize}
        \centering
        \includegraphics[width=.9\columnwidth]{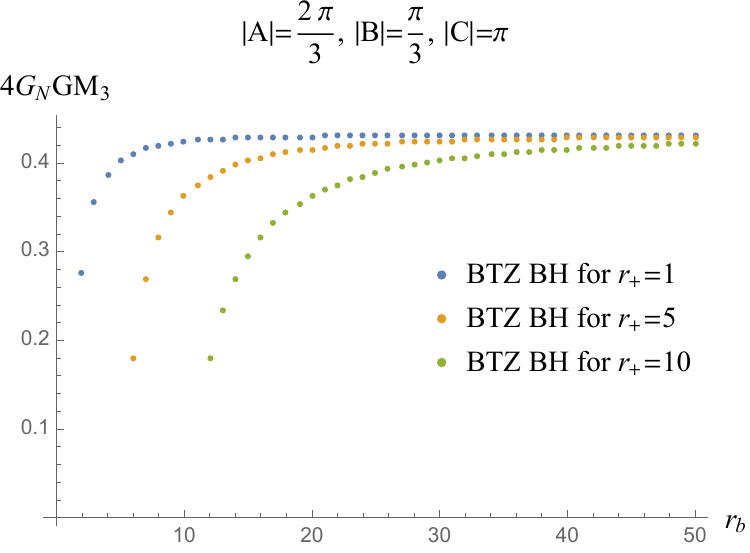}
    \end{minipage}
    \hspace{5mm}
    \begin{minipage}[h]{0.45\hsize}
        \centering
        \includegraphics[width=.95\columnwidth]{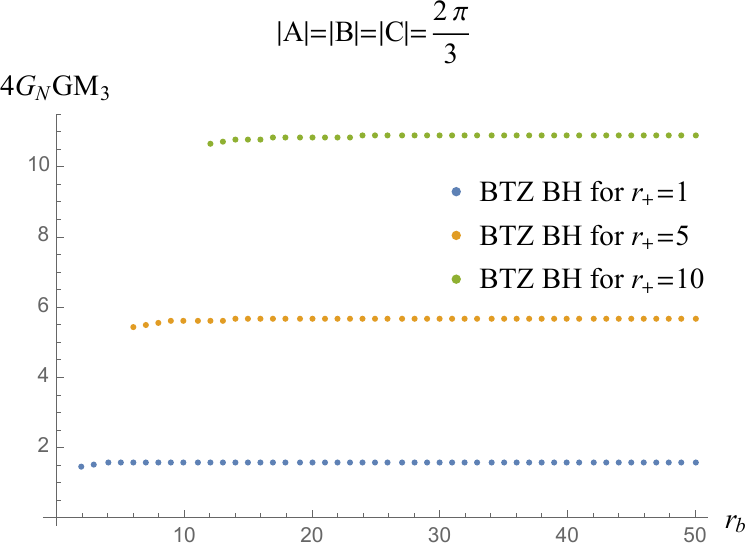}
    \end{minipage} \\
\caption{Plots showing finite-$r_b$ effects for the BTZ black hole up to around $r_b = r_+$.
Before (left panel) and after (right panel) one of the subsystems becomes larger than half of the total system, the magnitude of ${\rm GM}^{(3)}$ changes significantly even at finite $r_b$.}
\label{fig:finiter_BTZ}
        \centering\vspace{1cm}
\includegraphics[width=9cm]{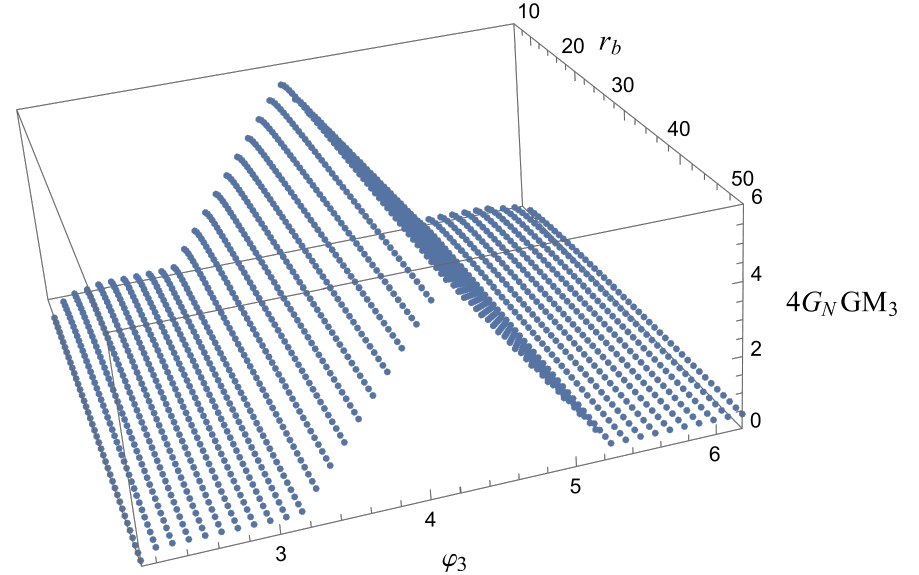}
    \caption{Full $r_b$ and $\varphi_3$-dependence of ${\rm GM}^{(3)}$ for static BTZ black hole with $r_+=5$.}
    \label{fig:finiter_BTZfull}
\end{figure}
One can also see finite-$r_b$ behavior by solving \eqref{eq:minimal} numerically. First, as in the discussion of Section~\ref{sec:BTZ}, the genuine multi-entropy ${\rm GM}^{(3)}$ exhibits multiple phases. Once the size of one subsystem exceeds one half of the total system, the volume-law–like contribution disappears, leaving only a subleading term of the same order as in global AdS${}_3$ (see the left panel of Figure \ref{fig:finiter_BTZ}). This remaining term decreases monotonically as the cutoff $r_b$ is lowered and shows a size dependence analogous to that in AdS${}_3$.
By contrast, in phases where the volume-law–like contribution is present, the reduction of ${\rm GM}^{(3)}$ with decreasing $r_b$ is much milder (see the right panel of Figure \ref{fig:finiter_BTZ}). This behavior suggests that, even at a moderate distance from the black hole, the structure of multipartite entanglement is significantly modified by backreaction effects on the spacetime geometry.
More detailed size dependence can be found in Figure \ref{fig:finiter_BTZfull}. Since the numerical analysis requires high precision near the horizon, we
restrict our computation to values of $r_b$ sufficiently close to the horizon $r = r_+$, but not exactly at it. 

\newpage
\section{Discussion}\label{sec:discussion}
In this paper, we investigated multipartite entanglement in holographic systems with a particular focus on AdS$_3$/CFT$_2$ in the presence of black holes, using multi-entropy as a quantitative probe. Our analysis reveals several qualitative and quantitative distinctions between vacuum AdS and black hole geometries, highlighting the role of black holes as efficient generators of multipartite entanglement.

\subsection*{Connection to tripartite Haar random state}
A closely related viewpoint on multipartite structure in holography was recently developed in \cite{Li:2025nxv}, where it was shown that a typical tripartite Haar random state contains essentially no distillable bipartite EPR-like entanglement between two subsystems, as long as each subsystem is smaller than half of the total system. Importantly, this holds even when the entanglement wedge of the two subsystems is connected, indicating that the connectivity of the entanglement wedge does not necessarily imply the presence of bipartite EPR pairs.

Our results are naturally compatible with this observation. The sharp transition we observe -- where genuine tripartite entanglement vanishes, up to the universal constant term, once one subsystem exceeds half of the total system  -- closely mirrors the half-system threshold identified in \cite{Li:2025nxv}. This parallel suggests that black-hole microstates,
indistinguishable from Haar-random states in many senses, predominantly store information in non-EPR-like, genuinely multipartite entanglement rather than in bipartite entanglement resources.

\subsection*{Area law of multi-entropy in vacuum AdS${}_3$}
While the main scope of this paper was on black holes, we also found an intriguing contribution to multi-entropy, 
\begin{align}
   S^{(3)} \text{or}\: {\rm GM}^{(3)}(A:B:C) \supset (\text{\# of boundary for subsystems})\times\dfrac{c}{6} \log \left[ \frac{2}{\sqrt{3}}\right],
\end{align}
where we translate the bulk results into CFT language using the Brown-Henneaux relation $c=\frac{3\ell}{2G_N}$~\cite{Brown:1986nw}. This contribution appears universally both in AdS${}_3$ and in the BTZ black hole. We therefore expect that it originates from the UV structure of 2D CFTs.
Since our analysis is restricted to a limited set of examples, more extensive studies are required to test this conjecture.

\subsection*{Emergence of Haar random structure well inside of black hole}
Motivated by the $T\overline{T}$-deformation and finite-cutoff AdS/CFT, we studied the finite-$r_b$ behavior of multi-entropy.
In this paper, we have focused only on $r_b > r_+$, outside the black hole.

On the other hand, recent developments suggest intriguing generalizations of this subject beyond the horizon~\cite{Araujo-Regado:2022gvw,Soni:2024aop,AliAhmad:2025kki}. In particular, if we place a holographic screen on the so-called special extremal slice~\cite{Hartman:2013qma} inside of the black holes ($r_b = 0$ for the BTZ black hole and $0<r_b<r_{\text{horizon}}$ for higher-dimensional black holes), and regard the boundary $t$-direction as a spatial coordinate, one can show that the dual state should be an absolutely maximally entangled (AME) state, provided that we impose Dirichlet boundary conditions on the screen~\cite{Anegawa:2025tio}.
The previous study only identified the bipartite entanglement structure, and it was not clear whether the AME state is closer to a stabilizer state or a Haar-random state.
Since the holographic calculation is effectively that of a BTZ black brane in the large-$r_+$ limit, we expect that the AME state inside the BTZ black hole also exhibits a multipartite entanglement structure similar to that outside, as discussed in the present paper.
This implies that the interior AME states are indeed well approximated by Haar-random states as well, thereby supporting recent assumptions in non-isometric code models~\cite{Akers:2022qdl}.
We leave a more detailed investigation of this perspective for future work.
Since certain constructions of the interior involve non-Hermitian holographic
theories~\cite{Araujo-Regado:2022gvw}, an extension of multi-entropy to such systems may also be required in line with the notion of pseudo entropy~\cite{Nakata:2020luh}.

\section*{Acknowledgement}
We thank Mitsuhiro Nishida for fruitful comments and discussions. We also thank Jonathan Harper and Norihiro Iizuka for useful conversations. T.A.~is supported by JSPS KAKENHI Grant No.~24K22886. K.T.~is supported by MEXT KAKENHI Grant No.~24H00972.  
\appendix
\section{Steiner trees with two vertices}\label{app:twov}
In this appendix, we extend our arguments in Section~\ref{sec:emb} to two-vertices. See Figure~\ref{fig:bh_four} for an illustration of the setup.

The area functional is given by
\begin{align}
L(Y_i,\Lambda_{j})&=\sigma(X_1,Y_1)+\sigma(X_2,Y_1)+\sigma(Y_1,Y_2)+\sigma(X_3,Y_2)+\sigma(X_1,Y_2)\nonumber\\
&+\Lambda_1(Y_1^2+1)+\Lambda_2(Y_2^2+1).
\end{align}
The equation of motion for $Y_i$ gives 
\begin{align}
\dfrac{-X_{1A}}{\sqrt{(-X_1\cdot Y_1)^2-1}}+\dfrac{-X_{2A}}{\sqrt{(-X_2\cdot Y_1)^2-1}}+\dfrac{-Y_{2A}}{\sqrt{(-Y_1\cdot Y_2)^2-1}}&=-2\Lambda_1Y_{1A}\\
\dfrac{-X_{3A}}{\sqrt{(-X_3\cdot Y_2)^2-1}}+\dfrac{-X_{4A}}{\sqrt{(-X_4\cdot Y_2)^2-1}}+\dfrac{-Y_{1A}}{\sqrt{(-Y_1\cdot Y_2)^2-1}}&=-2\Lambda_2Y_{2A}
\end{align}
together with 
\begin{align}
    Y_i^2+1=0
\end{align}
from the Lagrange multipliers. From these equations, one can solve $\Lambda_i$s as
\begin{align}
    \Lambda_1&=\dfrac{1}{2}\left[\dfrac{(-X_{1}\cdot Y_1)}{\sqrt{(-X_1\cdot Y_1)^2-1}}+\dfrac{(-X_{2}\cdot Y_1)}{\sqrt{(-X_2\cdot Y_1)^2-1}}+\dfrac{(-Y_1\cdot Y_{2})}{\sqrt{(-Y_1\cdot Y_2)^2-1}}\right]\\
    \Lambda_2&=\dfrac{1}{2}\left[\dfrac{(-X_{3}\cdot Y_2)}{\sqrt{(-X_3\cdot Y_2)^2-1}}+\dfrac{(-X_{4}\cdot Y_2)}{\sqrt{(-X_4\cdot Y_2)^2-1}}+\dfrac{(-Y_1\cdot Y_{2})}{\sqrt{(-Y_1\cdot Y_2)^2-1}}\right]
\end{align}
In what follows, we focus on the symmetric solution for simplicity. In this case, we have
\begin{align}
    -X_{1}\cdot X_{2}&=-X_{3}\cdot X_{4},\\
    -X_{1}\cdot X_{4}&=-X_{2}\cdot X_{3},\\
    -X_{1}\cdot Y_{1}&=-Y_{2}\cdot X_{4},\\
    -X_{2}\cdot Y_{1}&=-Y_{2}\cdot X_{3}.
\end{align}
Note that these conditions are valid even for BTZ black holes. For the empty AdS${}_3$, we have further constraints. 
From this, we immediately have
\begin{align}
\Lambda_1=\Lambda_2\equiv\Lambda.
\end{align}
Taking the inner product with $X_{1,2}, Y_{1,2}$ and using the above symmetric conditions, we obtain five independent equations of motion:
\begin{align}
\dfrac{1}{\sqrt{(-X_1\cdot Y_1)^2-1}}+\dfrac{(-X_1\cdot X_2)}{\sqrt{(-X_2\cdot Y_1)^2-1}}+\dfrac{(-X_1\cdot Y_2)}{\sqrt{(-Y_1\cdot Y_2)^2-1}}&=2\Lambda(-X_1\cdot Y_1),\\
\dfrac{(-X_1\cdot X_2)}{\sqrt{(-X_1\cdot Y_1)^2-1}}+\dfrac{1}{\sqrt{(-X_2\cdot Y_1)^2-1}}+\dfrac{(-X_{2}\cdot Y_2)}{\sqrt{(-Y_1\cdot Y_2)^2-1}}&=2\Lambda(-X_2\cdot Y_1),\\
\dfrac{(-X_1\cdot Y_2)}{\sqrt{(-X_1\cdot Y_1)^2-1}}+\dfrac{(-X_2\cdot Y_2)}{\sqrt{(-X_2\cdot Y_1)^2-1}}+\dfrac{1}{\sqrt{(-Y_1\cdot Y_2)^2-1}}&=2\Lambda(-Y_1\cdot Y_2),\\
\dfrac{(-X_1\cdot X_3)}{\sqrt{(-X_2\cdot Y_1)^2-1}}+\dfrac{(-X_1\cdot X_4)}{\sqrt{(-X_1\cdot Y_1)^2-1}}+\dfrac{(-X_1\cdot Y_1)}{\sqrt{(-Y_1\cdot Y_2)^2-1}}&=2\Lambda(-X_1\cdot Y_2),\\
\dfrac{(-X_2\cdot X_3)}{\sqrt{(-X_2\cdot Y_1)^2-1}}+\dfrac{(-X_2\cdot X_4)}{\sqrt{(-X_1\cdot Y_1)^2-1}}+\dfrac{(-X_2\cdot Y_1)}{\sqrt{(-Y_1\cdot Y_2)^2-1}}&=2\Lambda(-X_2\cdot Y_2),
\end{align}
where
\begin{align}
    2\Lambda=\dfrac{(-X_{1}\cdot Y_1)}{\sqrt{(-X_1\cdot Y_1)^2-1}}+\dfrac{(-X_{2}\cdot Y_1)}{\sqrt{(-X_2\cdot Y_1)^2-1}}+\dfrac{(-Y_1\cdot Y_{2})}{\sqrt{(-Y_1\cdot Y_2)^2-1}}
\end{align}
If we focus on empty AdS${}_3$ (we have additional constraints $-X_1\cdot Y_1=-X_2\cdot Y_1$ and $-X_2\cdot Y_2=-X_1\cdot Y_2$), these equations will further reduce to three independent ones. 
Note that these equations also become a bit simpler forms if we focus on $r_b\rightarrow\infty$ limit.

\section{Small $r_b$-expansion for AdS${}_3$}\label{app:smallr}
In this appendix, we compute the holographic tripartite multi-entropy in global AdS${}_3$ spacetime for \(r_b\ll1\). The inner product between $X_i$ and $X_j$ is given by
\begin{align}
    -X_i\cdot X_j&=\cos{(t_i-t_j)}+\left(\cos{(t_i-t_j)}-\cos{(\varphi_i-\varphi_j)}\right)r_b^2.
\end{align}
Note that these inner products are automatically expanded by $r_b$.  
We also have inner product between the boundary point $X_i$ and the bulk vertex $Y$,
\begin{align}
    -X_i\cdot Y&=\sqrt{1+r^2_{*}}\cos{(t_i-t_{*})}-r_{*}\cos{(\varphi_i-\varphi_{*})}r_b+\frac{1}{2}\sqrt{1+r^2_{*}}\cos{(t_i-t_{*})}r^2_b+\mathcal{O}(r^3_b).
\end{align}
%Then, the equations we need to solve are expressed as 
%\begin{align}
%x\left(\frac{x}{\sqrt{x^2-1}}+\frac{y}{\sqrt{y^2-1}}+\frac{z}{\sqrt{z^2-1}}\right)&=\frac{1}{\sqrt{x^2-1}}+\frac{a}{\sqrt{y^2-1}}+\frac{b}{\sqrt{z^2-1}}\\
%    y\left(\frac{x}{\sqrt{x^2-1}}+\frac{y}{\sqrt{y^2-1}}+\frac{z}{\sqrt{z^2-1}}\right)&=\frac{a}{\sqrt{x^2-1}}+\frac{1}{\sqrt{y^2-1}}+\frac{c}{\sqrt{z^2-1}}\\
%    z\left(\frac{x}{\sqrt{x^2-1}}+\frac{y}{\sqrt{y^2-1}}+\frac{z}{\sqrt{z^2-1}}\right)&=\frac{b}{\sqrt{x^2-1}}+\frac{c}{\sqrt{y^2-1}}+\frac{1}{\sqrt{z^2-1}}.
%\end{align}
%where
%\begin{align}
%    a&=a^{(0)}+a^{(1)}r^2\\
%    b&=b^{(0)}+b^{(1)}r^2\\
%    c&=c^{(0)}+c^{(1)}r^2
%\end{align}
%and
%\begin{align}
%    a^{(0)}&=\cos{(t_1-t_2)}\\
%    b^{(0)}&=\cos{(t_1-t_3)}\\
%    c^{(0)}&=\cos{(t_2-t_3)}\\
%    a^{(1)}&=\left(\cos{(t_1-t_2)}-\cos{(\phi_1-\phi_2)}\right)\\
%    b^{(1)}&=\left(\cos{(t_1-t_3)}-\cos{(\phi_1-\phi_3)}\right)\\
%    c^{(1)}&=\left(\cos{(t_2-t_3)}-\cos{(\phi_2-\phi_3)}\right).
%\end{align}

%Similarly, we define 
%\begin{align}
%    x&\equiv x^{(0)}+x^{(1)}r+x^{(2)}r^2+\mathcal{O}(r^3)\\
%    y&\equiv y^{(0)}+y^{(1)}r+y^{(2)}r^2+\mathcal{O}(r^3)\\
%    z&\equiv z^{(0)}+z^{(1)}r+z^{(2)}r^2+\mathcal{O}(r^3).
%\end{align}

For simplicity, let us assume the symmetric case, $-X_1\cdot X_2=-X_2\cdot X_3=-X_3\cdot X_1$. 
%\begin{align}
%    \frac{3x^2}{\sqrt{x^2-1}}&=\frac{2a+1}{\sqrt{x^2-1}}\\
%    x&=\sqrt{\frac{1+2(a^{(0)}+a^{(1)}r^2)}{3}}\label{eq:allsym}.
%\end{align}
%For \(t\) is constant, we can determine the value of \(a\), 
%\begin{align}
%    a^{(0)}&=1 \\
%    a^{(1)}&=\frac{3}{2}.
%\end{align}
Consequently, the solution of the equations is given by
\begin{align}
    -X_i\cdot Y&=\sqrt{\frac{1+(-2X_1\cdot X_2)}{3}}=1+\frac{1}{2}r^2_b+\mathcal{O}(r^3_b).
\end{align}
This leads
\begin{align}
    S^{(3)}(A:B:C)&=\frac{1}{4G_N}\cdot3\cosh^{-1}{\left(1+\frac{1}{2}r^2_b\right)}=\frac{3}{4G_N}r_b+\mathcal{O}(r^2_b).
\end{align}
Furthermore, since \(S(A)=S(B)=S(C)\) for the symmetric case, 
\begin{align}
    S(A)+S(B)+S(C)&=\frac{3}{4G_N}\cosh^{-1}{\left(1+\frac{3}{2}r^2_b\right)}=\frac{3}{4G_N}\sqrt{3}r_b+\mathcal{O}(r^2_b),
\end{align}
we obtain the genuine tripartite multi-entropy as
\begin{align}
    {\rm GM}^{(3)}(A:B:C)&=S^{(3)}(A:B:C)-\frac{1}{2}\left(S(A)+S(B)+S(C)\right)\\
    &=\frac{3}{4G_N}\left(1-\frac{\sqrt{3}}{2}\right)r_b+\mathcal{O}(r^2_b).
\end{align}
\bibliographystyle{JHEP}
\bibliography{reference.bib}
\end{document}